\documentclass[conference]{IEEEtran}

\IEEEoverridecommandlockouts
\usepackage{cite}
\usepackage{amsmath,amssymb,amsfonts}
\usepackage[linesnumbered,ruled,vlined]{algorithm2e}
\usepackage{setspace}
\usepackage{algorithmic}
\usepackage{graphicx}
\usepackage{textcomp}
\usepackage{xcolor}
\usepackage{enumerate}
\usepackage{ntheorem}
\usepackage{subfigure}
\usepackage{balance}
\usepackage{bm}
\usepackage{hyperref}
\newtheorem{definition}{Definition}
\newtheorem{lemma}{Lemma}

\usepackage{booktabs}
\usepackage{multirow}
\usepackage{array}
\usepackage{subfigure}
\usepackage{makecell}
\usepackage{tabularx}
\usepackage{amsfonts}
\usepackage{breqn}
\newcommand{\nop}[1]{}

\newtheorem*{proof}{Proof:}
\def\BibTeX{{\rm B\kern-.05em{\sc i\kern-.025em b}\kern-.08em
    T\kern-.1667em\lower.7ex\hbox{E}\kern-.125emX}}
\begin{document}

\makeatletter
\newcommand{\linebreakand}{%
  \end{@IEEEauthorhalign}
  \hfill\mbox{}\par
  \mbox{}\hfill\begin{@IEEEauthorhalign}
}
\makeatother

\title{\huge GCLS$^2$: Towards Efficient Community Detection Using Graph Contrastive Learning with Structure Semantics}



\author{
\IEEEauthorblockN{
Qi Wen\IEEEauthorrefmark{1},
Yiyang Zhang\IEEEauthorrefmark{1},
Yutong Ye\IEEEauthorrefmark{1},
Yingbo Zhou\IEEEauthorrefmark{1}
Nan Zhang\IEEEauthorrefmark{1}
Xiang Lian\IEEEauthorrefmark{2}, and
Mingsong Chen\IEEEauthorrefmark{1}}
\IEEEauthorblockA{\IEEEauthorrefmark{1}MoE Engineering Research Center on HW/SW Co-design and Application, East China Normal University, Shanghai, China}
\IEEEauthorblockA{\IEEEauthorrefmark{2}Department of Computer Science, Kent State University, Kent, OH 44242, USA}
\IEEEauthorblockA{\{51265902057, 51275902065, 52205902007, 52215902009, 51255902058\}@stu.ecnu.edu.cn}
\IEEEauthorblockA{xlian@kent.edu, mschen@sei.ecnu.edu.cn}}

\maketitle

\begin{abstract}
Due to the power of learning representations from unlabeled graphs, \textit{graph contrastive learning} (GCL) has shown excellent performance in community detection tasks.
Existing GCL-based methods on the community detection usually focused on learning attribute representations of individual nodes, which, however, ignores structural semantics of communities (e.g., nodes in the same community should be structurally cohesive). 
Therefore, in this paper, we consider the community detection under the community structure semantics and propose an effective framework for \textit{graph contrastive learning under structure semantics} (GCLS$^2$) to detect communities. To seamlessly integrate interior dense and exterior sparse characteristics of communities with our contrastive learning strategy, we employ classic community structures to extract high-level structural views and design a structure semantic expression module to augment the original structural feature representation.
Moreover, we formulate the structure contrastive loss to optimize the feature representation of nodes, which can better capture the topology of communities.
To adapt to large-scale networks, we design a \textit{high-level graph partitioning} (HGP) algorithm that minimizes the community detection loss for GCLS$^2$ online training. It is worth noting that we prove a lower bound on the training of GCLS$^2$ from the perspective of the information theory, explaining why GCLS$^2$ can learn a more accurate representation of the structure.
Extensive experiments have been conducted on various real-world graph datasets and confirmed that GCLS$^2$ outperforms nine state-of-the-art methods, in terms of the accuracy, modularity, and efficiency of detecting communities.

\end{abstract}



\section{Introduction}

\begin{figure}[t]
    \centering
    \subfigure[GCL-based community detection]{
        \includegraphics[width=0.95\linewidth]{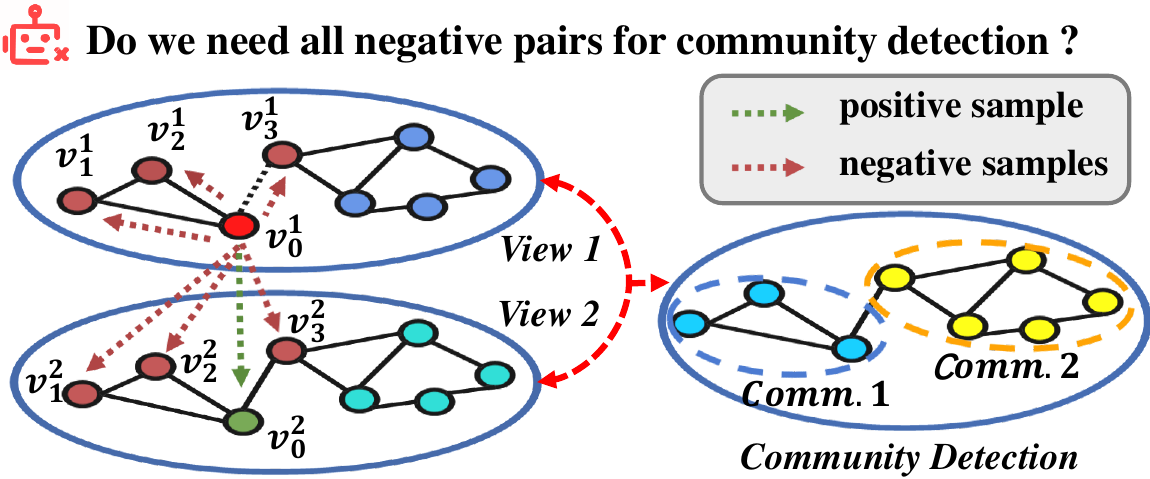}
        \label{subfig:fig_a}
    }\quad
    \subfigure[similarity matrix heatmap]{
        \includegraphics[width=1\linewidth]{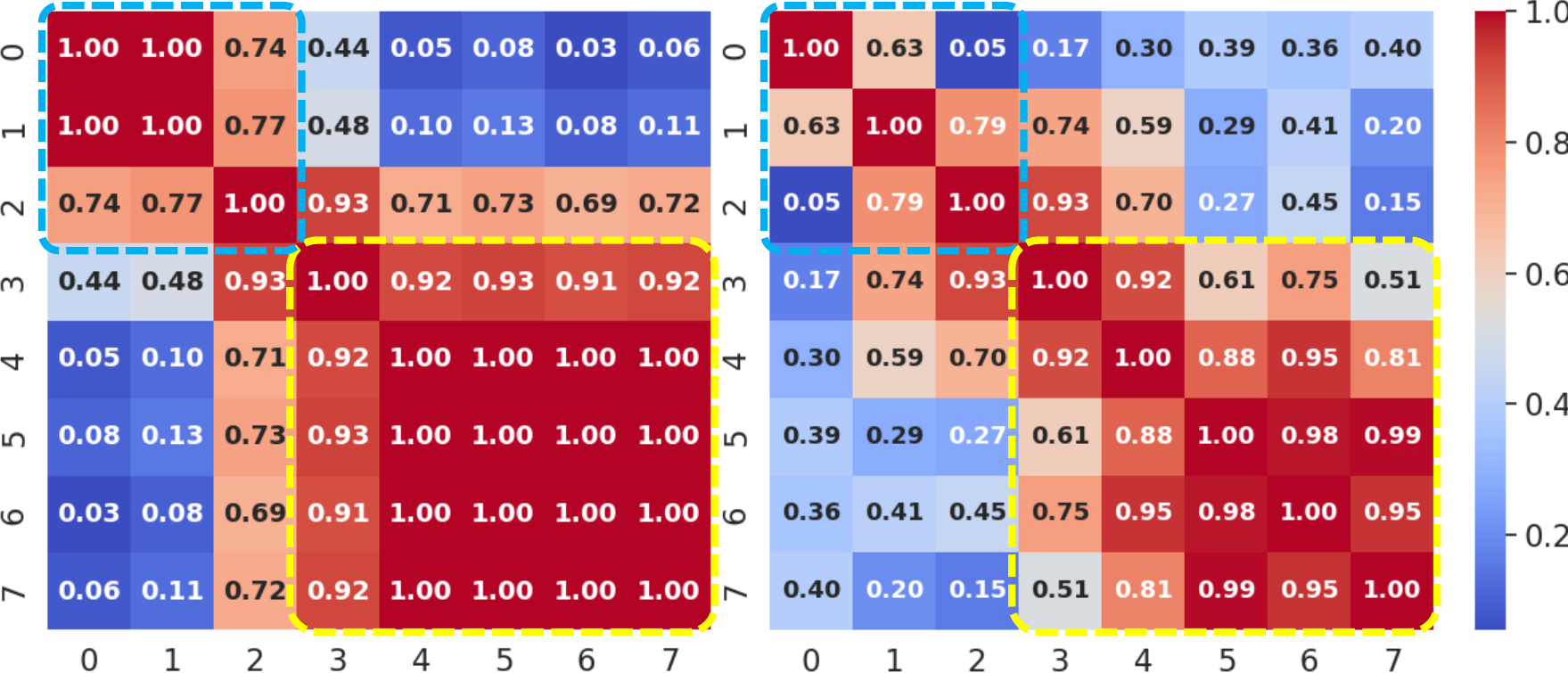}
        \label{subfig:fig_b}
    }
    \vspace{-2ex}
    \caption{A motivation example of graph structure semantics contrastive learning on the community detection.}
    \label{fig:motivation}
    \vspace{-3ex}
\end{figure}

The \textit{community detection} (CD) is one of fundamental and important problems in graph data analysis, and it plays an important role in many real-world applications such as the protein function prediction~\cite{whisstock2003prediction}, social network analysis~\cite{bedi2016community, LiuX0ZHPNYY20}, anomaly detection~\cite{keyvanpour2020ad}, and many others. Although traditional CD methods (e.g., hierarchical clustering~\cite{zarandi2018community}, spectral clustering~\cite{amini2013pseudo}, optimization-based methods~\cite{li2016multi}) can achieve high accuracy, they often incur high time costs and/or have poor scalability issue for conducting CD over large-scale data graphs, which limits these methods from being widely used in practical applications.

On the other hand, the supervised learning methods~\cite{ChenLB19, xin2017deep} have been used for CD tasks by mining from graph data, which take data graphs as the input through an end-to-end graph deep learning model, and output community prediction results via the trained neural networks. However, such supervised learning methods rely heavily on manually labeled data and have poor generalization performance due to the overfitting.

Recently, with the success of unsupervised contrastive learning in image and text representation learning, the \textit{graph contrastive learning} (GCL) paradigms (e.g., InfoNCE~\cite{he2020momentum}, GRACE~\cite{zhu2020deep}, GCA~\cite{zhu2021graph}, etc.) have been widely adopted in various tasks over graphs such as the link prediction and node classification, due to their excellent learning capability of the node information. 
However, these GCL-based methods do not consider structural relationships among nodes in communities,  which incurs a reduction in the accuracy and modularity of the detected communities.




\noindent {\bf Motivation:} Figure~\ref{subfig:fig_a} shows a traditional GCL-based community detection framework (i.e., GRACE~\cite{zhu2020deep}) with two communities (i.e., blue and gold groups), which first obtains negative views via data augmentation methods (e.g., edge perturbation, attribute mask, etc.), then constructs positive pairs (e.g., different views of the same nodes) and negative pairs (e.g., different views of neighbor nodes), and finally trains a model to minimize vector distances between positive pairs, while maximizing vector distances between negative pairs.
However, intuitively, the feature vector distances of nodes in the blue community (same as the gold community) should be close to each other, while the traditional GCL pulls them farther apart. 
Figure~\ref{subfig:fig_b} shows the heatmaps of the node similarity matrix from a 2-layer DNN applied to the data graph in Figure~\ref{subfig:fig_a}, comparing to the results without (left) and with (right) traditional GCL.
Note that the similarity of each pair of nodes is computed based on the dot product of the embedded representations of the nodes.
From Figure~\ref{subfig:fig_b}, we observe that the nodes expected to belong to the same community (e.g., those within blue and gold boxes) become less similar after GCL training, which will lead to low accuracy and modularity in the community detection.
This is because the embedding representations of nodes in the same community should be close to each other, while those from different communities should be distant, rather than having all nodes in the community being scattered in the GCL embedding space.

To tackle the aforementioned community detection problem, in this paper, we propose a \textit{graph contrastive learning} (GCL) based framework under structure semantics (named GCLS$^2$) for the community detection. Specifically, we start from the community structure and apply structure semantic contrastive learning to obtain better feature node representations by extracting structure semantics of graph data.
We construct a graph preprocessing module, using community dense structures (e.g., $k$-core~\cite{kong2019k}, $k$-truss~\cite{huang2014querying}, and $k$-clique~\cite{gregori2012parallel}) to obtain high-level and original structure views. 
For different views, we use a structure similarity semantic encoder to obtain structure features of the graph and design structure contrastive learning to fine-tune node structure feature representations of graph data, specific for the community detection task.

Note that, to address the issue that the memory size of local machine limits the information propagation in contrastive learning on large-scale graphs, we design an efficient \textit{high-level graph partitioning} (HGP) algorithm, which minimizes the loss of potential community integrity while ensuring balanced partitions. Furthermore, from the perspective of the information theory, we demonstrate the effectiveness of our proposed contrastive loss during the training. 

\vspace{1ex} \noindent {\bf Contributions:} The main contributions of this paper are summarized as follows:
\begin{itemize}
    \item We analyze the limitations of traditional GCL methods for the community detection task and explore a structure-driven approach to obtain better representation features.
    \item We propose a \textit{Graph Contrastive Learning with Structure Semantics} (GCLS$^2$) framework that can effectively extract structural representations of data graphs for the community detection.
    \item We design a \textit{Structure Similarity Semantic} (SSS) expression module in our proposed GCLS$^2$ framework to enhance the feature representation of structures.
    \item We devise an efficient \textit{high-level graph partitioning} (HGP) algorithm to effectively enable the training for contrastive learning over large-scale graphs.
    \item We provide a formal proof about the effectiveness of the proposed structure contrastive loss via information theory.
    \item Through extensive experiments, we confirm that our proposed GCLS$^2$ approach outperforms the supervised and unsupervised learning baselines on the accuracy, modularity, and efficiency of the community detection.
\end{itemize}

The remainder of this paper is organized as follows. 
Section~\ref{sec:PD} formally defines the \textit{community detection} problem in graphs. 
Section~\ref{sec:MT} proposes our GCLS$^2$ model and HGP algorithm, and proves the effectiveness of our GCLS$^2$ model training.
Section~\ref{sec:EX} conducts comprehensive experiments to compare our GCLS$^2$ approach with eight state-of-the-art algorithms. 
Section~\ref{sec:RW} introduces related works on the \textit{community detection} and \textit{graph contrastive learning}.
Finally, Section~\ref{sec:CC} concludes this paper.


\section{Problem Definition}
\label{sec:PD}

In this section, we give formal definitions of the graph data model and community detection problem.

\subsection{Graph Data Model}
We first provide the formal definition of a data graph $G$.

\begin{definition}
    \textbf{(Data Graph, \bm{$G$})} A data graph $G$ is in the form of a quintuple $(V(G), E(G), \Phi(G), A, X)$, where $V(G)$ is a set of nodes, $v_i$, $E(G)$ represents a set of edges $e(v_i, v_j)$ (connecting two ending nodes $v_i$ and $v_j$), $\Phi(G)$ is a mapping function: $V(G) \times V(G) \rightarrow E(G)$, and $A \in \mathbb{R}^{N \times N}$ and $X \in \mathbb{R}^{N \times F}$ are the adjacency and attribute matrices, respectively. Here, $A_{ij}=1$, if $(v_i,v_j) \in E(G)$, and $X$ contains attribute vectors, $x_i \in \mathbb{R}^F$, of $N$ vertices $v_i$.
    \label{def:data_graph}
\end{definition}

\begin{figure*}[!t]
    \centering
    \includegraphics[width=0.9\textwidth]{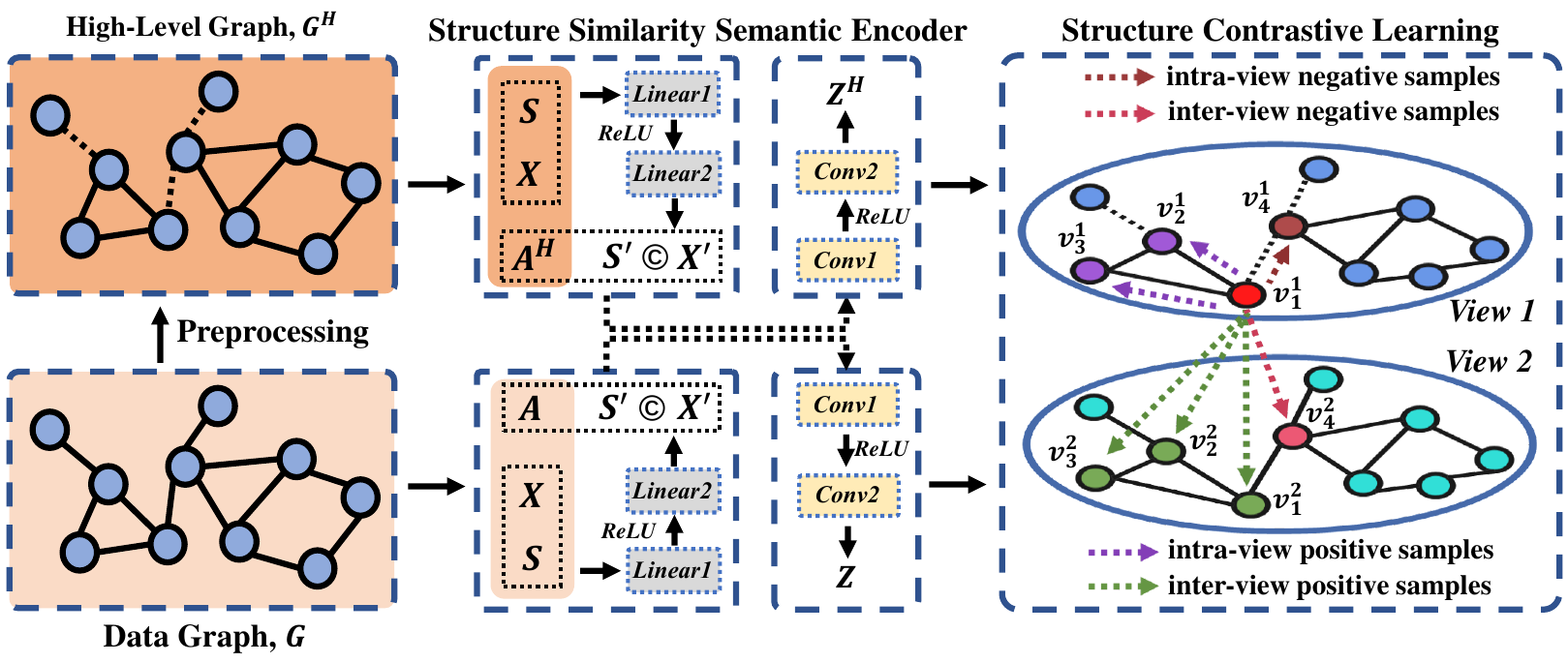}
    \vspace{-0.1in}
    \caption{Overview  of  our GCLS$^2$ framework. The {\normalsize{\textcircled{\scriptsize{C}}}\normalsize} denotes the concatenation of vectors.}
    \label{fig:framework}
\end{figure*}

Definition~\ref{def:data_graph} defines the data graph model, which is useful in many real applications such as social networks \cite{huang2014querying,zhang2023top}, transportation networks \cite{rai2023top}, and bioinformatic networks \cite{frith2020minimally,zhao2012rapsearch2}.

\subsection{The Community Detection}

The community detection is a fundamental task in the network analysis. The target is to partition the data graph into multiple subgraphs (i.e., communities) that are internally dense and externally sparse. The resulting communities can be used to uncover important structures and patterns within complex networks and extract valuable knowledge from network data in diverse domains.

Formally, we define the community detection as follows.

\begin{definition}
\textbf{(Community Detection, CD)} Given a data Graph $G$ and a positive integer parameter $L$, the \textit{community detection} (CD) problem obtains a set of disjoint subgraphs (i.e., communities), $g = \{g_1, g_2, \cdots, g_L\}$, in the data graph $G$, where each community $g_i$ ($1\leq i \leq L$) in $g$ contains classic community structures (e.g., $k$-core, $k$-truss, $k$-plex).
\label{def:CSS}
\end{definition}

In the literature of social networks, the commonly-used dense community semantics with high structural cohesiveness include $k$-core~\cite{yang2012defining}, $k$-truss~\cite{huang2014querying}, and $k$-plex~\cite{chang2022efficient}. Specifically, for $k$-core, the degree of each node in $g$ is greater than or equal to $k$; for $k$-truss, each edge is contained in at least $(k-2)$ triangles; for $k$-plex, the degree of each node in $g$ is greater than $(|V(g)|-k)$.

\section{Graph Contrastive Learning with Structure Semantics for the Community Detection}
\label{sec:MT}
In this section, we present our proposed GCLS$^2$ approach in detail. First, we will briefly discuss our framework. Then, we will detail the graph augmentation method for high-level structure and an encoder for extracting graph structure similarity semantics. After that, we will introduce how to design the structure contrastive loss.

\subsection{The GCLS$^2$ Framework}

Figure~\ref{fig:framework} shows the overall GCLS$^2$ framework for the community detection over a data graph $G$. We first use classical community structures to extract a high-level structure graph $G^H$ from the data graph $G$ for preprocessing, then we obtain the structure similarity semantic vectors of the data graph and the high-level structure graph through structural similarity matrix computation and semantic encoding, and we obtain the feature representations of the two graphs by GCN encoder. After that, we use the feature representations for the structure contrastive learning so that the features better match the structure semantic information of community detection.

\subsection{Graph Preprocessing}
\label{sec:gp}
In GCLS$^2$ model, the graph preprocessing is used to obtain the high-level graph that deeply expresses the structure information of the graph. 
We show an example of the graph preprocessing in Figure~\ref{fig:processing}.
For a data graph $G= (V(G), E(G), \Phi(G), A, X)$, we have the adjacency matrix $A \in \{0,1\}^{N \times N}$ and arrtibute matrix $X \in \mathbb{R}^{N \times F}$, where $N$ denotes the size of $V(G)$ and $F$ denotes the number of attributes. We consider using a set, $M = \{M_1, M_2, \cdots, M_m\}$, of the classical community structure patterns (i.e., $k$-core, $k$-truss, $k$-plex) to extract the high-level structure of $G$. 
Specifically, we count the substructures of the original graph using each pattern $M_i \in M$, and we initialize a dictionary $Dic$ to store the number of specific substructural patterns held for each edge pair $\left \langle v_i,v_j \right \rangle$. 
In Figure~\ref{fig:processing}, we use two substructures (triangle and pentagon, which are 1-core and 3-plex structures, respectively) for pattern counting, and the dashed lines in Figure~\ref{fig:processing} denote the edges with a pattern count of 0 and masked. 
In fact, other specified dense structures (e.g., rectangles) could also be used for the task. However, in this paper, we focus on pattern counting based on the two fundamental structures mentioned above.
In this way, we can obtain a high-level structure graph $G^H$ and a high-level adjacency matrix $A^H$. 
It is worth noting that we did not remove those nodes whose neighboring edges all have a pattern number of 0, and retained their attribute values for subsequent structure contrastive learning to maintain model smoothing.

After that, we extract the elements of the dictionary $Dic$ and construct a structure similarity matrix $S \in \mathbb{R}^{N \times N}$ for $G$ and $G^H$, the computation of each $s_{i,j} \in S$ as follows:
\begin{equation}
    s_{i,j} = \frac{Dic(v_i,v_j) + 1}{max(Dic(\cdot)) + 1},
    \label{eq:sij}
\end{equation}
where $max(Dic(\cdot))$ denotes the maximum value of the pattern count in the dictionary $Dic$. The $S$ represents high-level structure features in $G$ and $G^H$ that can deeply represent structure aggregation information.

\begin{figure}[!t]
    \centering
    \includegraphics[width=0.4\textwidth]{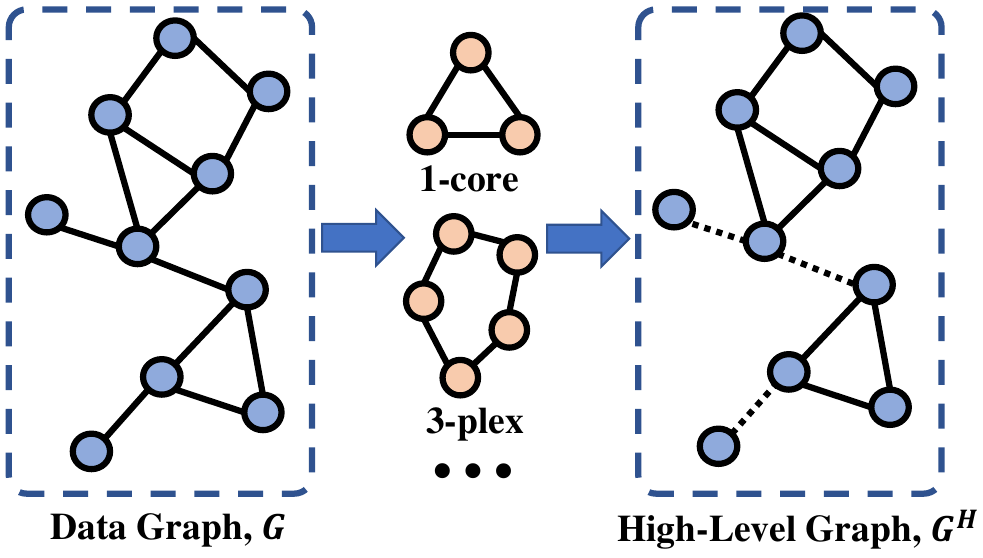}
    \vspace{-0.1in}
    \caption{An example of  graph preprocessing.}
    \label{fig:processing}
    \vspace{-3ex}
\end{figure}

\subsection{Structure Similarity Semantic Encoder}
Different from most of the previous graph encoders that directly convolve the attribute matrix $X$ and the adjacency matrix $A$ to extract graph features, we design a \textit{structure similarity semantic} (SSS) module to extract the semantic representations of structures and attributes for low-level features $S'$ and $X'$, which is to be able to better provide a more intuitive semantic representation for subsequent graph feature extraction. For the structure similarity matrix $S$ and attribute matrix $X$, we use a two-layer DNN $f: \mathbb{R}^{N \times N} \rightarrow \mathbb{R}^{N \times d}$ for semantic features $S'$ and $X'$, that is:
\begin{equation}
    S' = f(S) = W^{(2)}_{dnn}(\sigma(W^{(1)}_{dnn} S+b^{(1)}_{dnn}))+b^{(2)}_{dnn},
    \label{eq:s'}
\end{equation}
\begin{equation}
    X' = f(X) = W^{(2)}_{dnn}(\sigma(W^{(1)}_{dnn} X+b^{(1)}_{dnn}))+b^{(2)}_{dnn},
    \label{eq:x'}
\end{equation}
where $\sigma(\cdot)$ denotes the ReLU activation function, $W_{dnn}$ is the neural network trainable weight parameters, and $b_{dnn}$ is the bias. The (1) and (2) denote the 1-th layer and 2-th layer, respectively.   
The second term of the equation represents the process of message-passing in the DNN network. The output dimensions of both features are $d$.

Then, to encode the attribute semantic feature $X'$ and the structure similarity semantic feature $S'$ into a unified graph feature space without destroying the respective semantic representations, we aggregate the two features by concatenation and use them in the subsequent feature extraction.

Next, we learn a GCN encoder $\mathbb{F}: \mathbb{R}^{N \times N} \times \mathbb{R}^{N \times 2d} \rightarrow \mathbb{R}^{N \times L}$ to obtain the node representations of  $G$ and $G^H$, where $L$ denotes the output dimension of the encoder, both of which share weights during the learning process. The input of $G$ is $(A, AGG(S', X'))$ and the input of $G^H$ is $(A^H, AGG(S', X'))$, where the $AGG(\cdot)$ is concatenation aggregated function. So that we can obtain the node features $Z \in \mathbb{R}^{N \times L}$ of the graph by the generalized equation:
\begin{equation}
    Z = \mathbb{F}(A, AGG(S', X')) = \{z_i | v_i \in V(G)\}.
    \label{eq:Z}
\end{equation}
The detailed message-passing process of GCN is as follows:
\begin{equation}
    z_i^{(l+1)} = \sigma(\sum_{v_j \in N(v_i)} \hat{D}^{-\frac{1}{2}} \hat{A} \hat{D}^{-\frac{1}{2}} z_j^{(l)}W^{(l)}_{gcn}+b^{(l)}_{gcn}),
\end{equation}
where $l$ denotes the l-th layer, $\hat{A} = A + I$ denotes the message self-passage, $I$ denotes the unit matrix, $D$ denotes the degree matrix, $\sigma(\cdot)$ denotes the activation function, we use the ReLU in GCLS$^2$, the $W_{gcn}$ and $b_{gcn}$ denote the trainable weight parameters and bias in GCN and the $N(v_i)$ denotes the neighbors of $v_i$. The 0-th node features as follows:
\begin{equation}
    z_i^{(0)} = AGG(s_{i}^{'}, x_{i}^{'}) | s_{i}^{'} \in S', x_{i}^{'} \in X'.
\end{equation}

Similarly, for the high-level structure graph $G^H$, the representation of node features is as follows:
\begin{equation}
    Z^H = \mathbb{F}(A^H, AGG(S', X')) = \{z_i | v_i \in V(G^H)\}.
    \label{eq:ZH}
\end{equation}
The detailed message-passing is the same as the graph $G$.

\subsection{Structure Contrastive Learning}
Most existing GCL methods overlooked that nodes within a community should have similar representations, while inter-community nodes should differ. 
Therefore, we proposed a structure contrastive learning, which uses the graph's high-level structure to divide the positive and negative pairs of the GCL for structure learning, thereby enhancing the community representation of nodes.

\begin{figure*}[!t]
    \centering
    \includegraphics[width=0.85\linewidth]{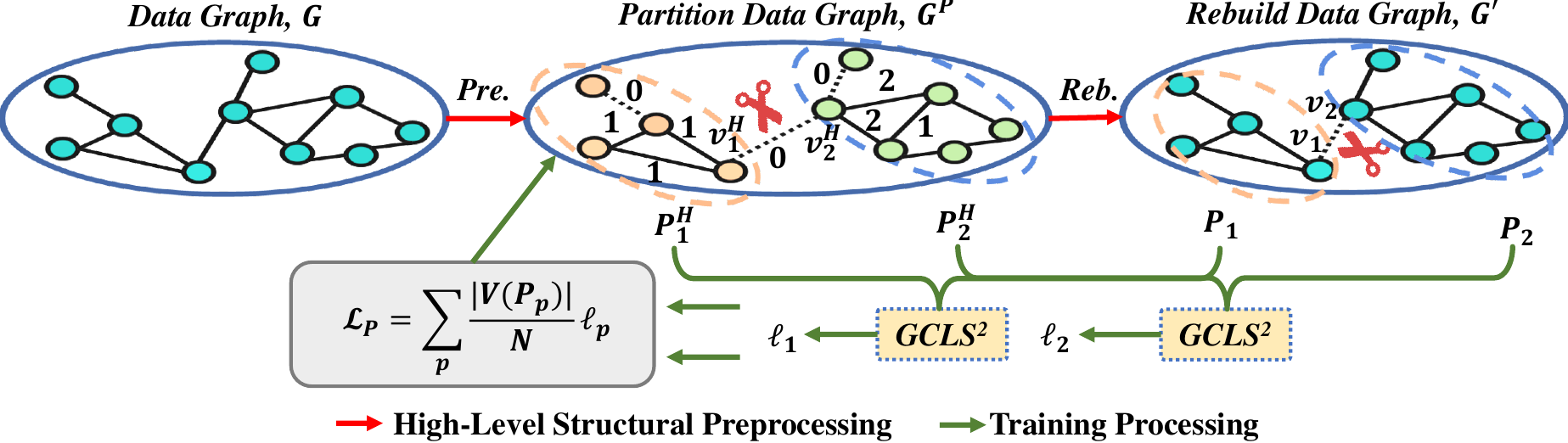}
    \caption{The framework of the HGP algorithm.}
    \label{fig:GCLS2H}\vspace{-3ex}
\end{figure*}

Since GCLS$^2$ does not use an activation function in the last layer of the GCN, we will do $L_2$-normalization on the learned node representations to ensure that the magnitude of the data is uniform before the structure contrastive learning.
Moreover, we take the original data graph $G$ and high-level structure graph $G^H$ as two views for structure contrastive learning, which is because the $G$ is unmodified and contains accurate and rich graph-based information, and the $G^H$ contains rich and dense community information.

Let $z_i^{1}$ and $z_i^{2}$ denote the $v_i$'s $L_2$-normalized embedding learned by view 1 and view 2, respectively. 
We consider $z_i^{1}$ as an anchor, where there are two positive sample pairs: \romannumeral1) intra-view high-level neighbors in the same dense structure, i.e., $\{z_j^{1} | v_j \in N(v_i)^H \}$; \romannumeral2) inter-view high-level neighbors in the same dense structure, i.e., $\{z_i^{2}, z_j^{2} | v_j \in N(v_i)^H \}$. Then, the number of positive sample pairs with anchor $z_i^{1}$ is $2|N(v_i)^H| + 1$, where $|N(v_i)^H|$ denotes the number of neighbors of $v_i$ in high-level structure graph $G^H$. With the rest formed two negative sample pairs: \romannumeral1) intra-view nodes not in the high-level neighbors, i.e., $\{z_j^{1} | v_j \notin N(v_i)^H \}$; \romannumeral2) inter-view not in the high-level neighbors, i.e., $\{z_i^{2}, z_j^{2} | v_j \notin N(v_i)^H \}$.
Finally, the structure contrastive loss of $z_i^{1}$ between view 1 and view 2 is formulated as:
\begin{equation}
    pos(z_i^{1}) = \sum\limits_{v_j \in N(v_i)^H}(e^{sim(z_i^1, z_j^1) / \tau} + e^{sim(z_i^1, z_j^2) / \tau}),
\end{equation}
\begin{align}
    neg(z_i^{1}) &= \sum\limits_{v_j \notin N(v_i)^H} (e^{sim(z_i^1, z_j^1)/ \tau}+e^{sim(z_i^1, z_j^2)/ \tau}),
\end{align}
\begin{equation}
    \ell(z_i^{1}) = -log \frac{pos(z_i^{1}) + e^{sim(z_i^1, z_i^2) / \tau}}{neg(z_i^{1})+ pos(z_i^{1}) + e^{sim(z_i^1, z_i^2) / \tau}},
    \label{loss:node}
\end{equation}
where the $sim(z_i,z_j) = z_i \cdot z_j$ denotes the vector similarity between $z_i$ and $z_j$, and  $\tau$ denotes a temperature parameter.  

Since the structure contrastive learning is computed at the embedding level of the nodes, to simplify the representation, we use $v_i$ instead of $z_i$ in the Structure Contrastive Learning of Figure~\ref{fig:framework} as an example.
We set $v_1^1$ as the anchor, then the positive sample pairs of $v_1^1$ have: \romannumeral1) the intra-view, \{$\left \langle v_1^1, v_2^1 \right \rangle$, $\left \langle v_1^1, v_3^1 \right \rangle$\}; \romannumeral2) the inter-view, \{$\left \langle v_1^1, v_1^2 \right \rangle$, $\left \langle v_1^1, v_2^2 \right \rangle$, $\left \langle v_1^1, v_3^2 \right \rangle$\}. The negative sample pairs of $v_1^1$ have: \romannumeral1) the intra-view, \{$\left \langle v_1^1, v_4^1 \right \rangle$, $\cdots$, $\left \langle v_1^1, v_n^1 \right \rangle$\}; \romannumeral2) the inter-view, \{$\left \langle v_1^1, v_4^2 \right \rangle$, $\cdots$, $\left \langle v_1^1, v_n^2 \right \rangle$\}.

Minimizing Eq.~\ref{loss:node} will maximize the agreement between pairs of positive samples and minimize the agreement of pairs of negative samples. 
Through the computational propagation of mutual information between nodes, the feature representations of each node will agree with the high-level structure pattern's feature representations within the other view itself and between the intra-view and inter-view.

Since the two views are symmetric, the node structure contrastive loss of the other view $\ell(z_i^{2})$ is defined similarly to Eq.~\ref{loss:node}. Then we minimize the loss mean of the two views to give the overall contrastive loss as follows:
\begin{equation}
    \mathcal{L} = \frac{1}{2N}\sum_{i=1}^N[\ell(z_i^{1})+\ell(z_i^{2})].
    \label{eq:loss}
\end{equation}

\subsection{The High-Level Graph Partitioning}

For large-scale graphs, the memory of most machines cannot keep up with the increase of the number of neighbors of a GNN, making many graph representation methods like graph contrastive learning difficult to execute on GPUs.

Existing graph partitioning algorithms for large-scale graph learning mainly have two categories, vertex partitioning~\cite{karypis1998fast, stanton2012streaming, tsourakakis2014fennel} and edge partitioning~\cite{xie2014distributed,petroni2015hdrf}.
Note that, the edge partitioning assigns edges to the corresponding subgraphs, leading to vertex redundancy and multiple representations of some vertices during the graph learning. This redundancy increases computational overhead for query processing. Thus, in this paper, we instead adopt a vertex partitioning approach, which assigns vertices to subgraphs in large-scale data graphs.

The two objectives of the graph partitioning are load balancing and minimizing the cost of cutting. Optimizing these two objectives together is called the \textit{Balanced Graph Partitioning} problem, which is NP-hard~\cite{zhang2017graph}.
Traditional vertex partitioning algorithm, METIS~\cite{karypis1998fast}, uses the sparsification and refinement aggregation of consecutive vertices and edges to partition the data graph, but requires random traversal and scaling of the entire graph structure, which tends to break the community structure, resulting in a large loss of boundary features. The LGD~\cite{stanton2012streaming} algorithm uses a greedy strategy to aggregate neighboring nodes in the same partition and ensures load balancing in each partition.
As shown in Figure~\ref{fig:motivation}(a), not all neighboring nodes belong to the same community. The partitions generated by the LDG algorithm can lead to abnormal community detection results.
Moreover, METIS and LDG perform poorly on large-scale data graphs even though they try to keep the load balanced as much as possible.

To adapt to our community detection task, we propose a framework for high-level graph partitioning (HGP) to minimize the damage to the community structure, which is illustrated in Figure~\ref{fig:GCLS2H}.

\noindent
\textbf{The HGP Framework:} Our HGP framework consists of three main steps: 
(1) high-level graph partitioning, 
(2) rebuilding of the original data graph, and
(3) partition training of the model.

Note that, cutting edges may break the community structure and affect the downstream detection results, but this is unavoidable.
Therefore, our goal of HGP is to minimize the loss of the community structure while ensuring load balancing in graph partitioning, defined as follows.


\begin{definition}
    \textbf{(High-Level Graph Partitioning, HGP)} Given a data graph $G(V,E)$, the number, $p$, of partitions, and the balanced partition rate $a$, the objectives of HGP are given by:
    \begin{equation}
    \begin{cases}
         \max_{i \in [1, p]}|V(part_i)|\leq \frac{(1+a)|V(G)|}{p} \\
         \min |E(G)-(\bigcup_{i \in [1,p]}E(part_i))|
    \end{cases},
    \label{eq:hgp}
    \end{equation}
    where $V(part_i)$ and $E(part_i)$ denote the sets of vertices and edges in partition $part_i$, respectively.
\end{definition}

To achieve this goal, we design an HGP optimization algorithm as shown in Algorithm~\ref{alg:hgp}. 
Specifically, we first obtain a high-level adjacency matrix $A^H$ precomputed in Section~\ref{sec:gp} for high-level structural partitioning of the data graph $G$ (line 1). 
Next, we make a judgment on each vertex $v_i \in V(G)$: if the current vertex $v_i$ has no neighbors in the high-level adjacency matrix $A^H$, i.e., this is a solitary vertex, 
then we consider assigning $v_i$ to the partition with the highest number of neighbors in the original data graph $G$ (lines 2-4). As shown in the \textit{partition data graph} $G^P$ in Figure~\ref{fig:GCLS2H}, this yellow solitary vertex in the upper left corner is assigned to partition $P^H_1$ because all its original neighbors are in partition $P^H_1$. 
Otherwise, we test each partition $j=1,\cdots,p$: if the partitioning size is less than the load limit, we query how many high-level neighbors of $v_i$ exist in the current partition $part_j$, we aim to divide $v_i$ into the partition with the highest number of high-level neighbors, so we compute the current partition score,
\begin{equation}
    s(v_i,part_j) = |part_j \bigcap N(v_i)^H| \cdot \frac{|V(G)|}{p(|part_j|+1)},
\end{equation}
where the second term is a load limit to limit the size of the partition (lines 5-9).
Finally, we choose to assign $v_i$ to the partition with the highest score $s$ (line 10) until we return all partitioning results for each vertex (line 11).

After completing the high-level graph partitioning, we perform the mapping partitioning of the original graph to rebuild the data graph.
Specifically, we synchronously map the vertex partition IDs obtained from the HGP algorithm into the original data graph, resulting in $p$ partitions with the same size as those obtained by the HGP algorithm.
As shown in Figure~\ref{fig:GCLS2H}, the $v_1^H$ and $v_2^H$ in $G^P$ are assigned to different partitions, and after mapping, the original data graph cuts the edge $e(v_1,v_2)$ to form the rebuild data graph $G'$.

Actually, we obtain $p$ high-level structural subgraphs and $p$ original subgraphs after the HGP partitioning and rebuild.
In the training process of contrastive learning, we take each pair of subgraphs (i.e., $\left \langle P_1^H,P_1 \right \rangle$) for contrastive training.
Since the size of the cut subgraphs can be controlled by the specified parameter $p$, users can completely adjust the parameter $p$ through the local memory for the online contrastive learning training. 
In Figure~\ref{fig:GCLS2H}, $\left \langle P_1^H,P_1 \right \rangle$ and  $\left \langle P_2^H,P_2 \right \rangle$ are learned by the same GCLS$^2$ model, which results in the corresponding loss values $\ell_1$ and $\ell_2$.
To tune the learning globally, we design the following global loss for backpropagation learning through these loss values.
\begin{equation}
    \mathcal{L}_P=\sum\limits_p\frac{|V(P_p)|}{N}\ell_p.
    \label{eq:L_p}
\end{equation}

It is worth noting that since the training is on different subgraph pairs and our partitioning results maximally guarantee the load balancing of each partition through the HGP algorithm.
Therefore, we can further improve the training efficiency of large-scale data graphs by parallel training.

\begin{algorithm}[!t]
    \caption{HGP Algorithm}
    \label{alg:hgp}
    \KwIn{
        \romannumeral1) the data graph $G$, and
        \romannumeral2) the number, $p$, of partitions}
    \KwOut{
        graph partitions in $pid$
    }

    $A^H \leftarrow A$ \hfill $\triangleright$ \CommentSty{obtain by Section~\ref{sec:gp}}

    \For{each vertex $v_i \in V(G)$}{

        \tcp{solitary vertex}
        
        \eIf{$N(v_i)^H = \emptyset$}{

            $pid[v_i] = \mathop{\arg\max}\limits_{j \in \{1,\cdots, p\}}\{|part_j \bigcap N(v_i)|\}$
            
        }{

            \For{each partitions $j=1,\cdots,p$}{

                \If{$|part_j| < \frac{|V(G)|}{p}$}{

                    \tcp{high-level neighbors in partition $j$}
                
                    $part_j \bigcap N(v_i)^H$ 
    
                    \tcp{partition score with load limit}
                    
                    $s(v_i, part_j) = |part_j \bigcap N(v_i)^H| \cdot \frac{|V(G)|}{p(|part_j|+1)}$
                    
                }

            }

            $pid[v_i] = \mathop{\arg\max}\limits_{j \in \{1,\cdots, p\}}\{s(v_i,part_j)\}$
        }
        
    }
    
    \Return $pid$
\end{algorithm}

        

\subsection{Model Training and Complexity Analysis}

\noindent
\textbf{The Model Training: }The detailed GCLS$^2$ training process is shown in Algorithm~\ref{alg:train}. 
Specifically, we first obtain the high-level structure graph $G^H$ and high-level adjacency matrix $A^H$ from the original data graph $G$, and construct a dictionary $Dic$ by counting the number of specified patterns for each edge of $G$, then extract the structure similarity matrix $S$ from the $Dic$ (lines 1-3). 
For each training iteration, we first use the SSS module to extract structural semantics representation $S'$ and attribute semantics representation $X'$ (lines 4-6). Next, we employ GCNs with weight shared to extract node features from both $G$ and $G^H$, resulting in the node feature matrix $Z$ and the high-level node feature matrix $Z^H$ (lines 7-8). 
Then, we calculate the similarity between positive samples and the similarity between negative samples to obtain the structure contrastive loss $(\ell(z_i^{1}), \ell(z_i^{2}))$ for each vertex $v_i$ (line 9).
After that, we update the GCLS$^2$ network parameters by minimizing all structure contrastive loss $\mathcal{L}$ on $Z$ and $Z^H$ (lines 10-11). 
Finally, after $\varepsilon$ training iterations, we return the node features $Z$ obtained for subsequent community detection (line 12).

\noindent
\textbf{The Complexity Analysis: } We analyze the time complexity and space complexity of the GCLS$^2$ model by the number of floating point operations (FLOPs) and space overhead required by the GCLS$^2$ model. 
Specifically, let the time required for each unit floating point operation be $O(1)$.
For the \textit{Structure Similarity Semantic Encoder} module, the time complexity of the SSS module is $O(Nd^2)$, and the time complexity of the node representation extractor is $O(|E|d)$.
For the \textit{Structure Contrastive Learning} module, the similarity of each pair of vertices can be realized by vector dot product, which would cost $O(d)$. Then, the time complexity of all structure contrastive loss is $O(Nd)$.
Therefore, the time complexity of the training process of GCLS$^2$ is $O(Nd^2+|E|d)$

For the \textit{Structure Similarity Semantic Encoder} module, the space complexity is $O(N^2+d^2)$, which contains the input graph size and the SSS parameter size. 
Similarly, the space complexity of the node representation extractor is $O(Nd+d^2)$.
For the \textit{Structure Contrastive Learning} module, the computation of the structure contrastive loss can be performed on the feature matrices $Z$ and $Z^H$ with a space complexity of $O(N^2)$.
Since the individual training process is serial and $d$ is significantly smaller than $N$, the space complexity of the training process of GCLS$^2$ is $O(N^2)$.

For the HGP algorithm, the time complexity of the partitioning is $O(pN+|E|)$. The space complexity of the partitioning is $O(N)$,

\begin{algorithm}[!t]
    \caption{GCLS$^2$ Training Algorithm}
    \label{alg:train}
    
    \KwIn{
        \romannumeral1) a data graph $G$ with adjacency matrix $A$ and attribute matrix $X$, and
        \romannumeral2) the number of training epochs $\varepsilon$
    }
    \KwOut{
        features $Z$
    }
    \setstretch{1.25}
    \tcp{high-level graph preprocessing}

    $G^H \gets G$, $A^H \gets A$ 

    construct a dictionary $Dic$ by counting the number of specified patterns for each edge of $G$
    
    $S \gets \sum_{i,j}\frac{Dic(v_i,v_j)+1}{\max (Dic(\cdot))+1}$  \hfill $\triangleright$ \CommentSty{Eq.~\ref{eq:sij}}

    
    \tcp{training process}
    
    \For{epoch $\gets 1,2,\cdots, \varepsilon$}{
        

        $S' \gets W^{(2)}_{dnn}(\sigma(W^{(1)}_{dnn} S+b^{(1)}_{dnn}))+b^{(2)}_{dnn}$ \hfill $\triangleright$ \CommentSty{Eq.~\ref{eq:s'}}


        $X' \gets  W^{(2)}_{dnn}(\sigma(W^{(1)}_{dnn} X+b^{(1)}_{dnn}))+b^{(2)}_{dnn}$ \hfill $\triangleright$ \CommentSty{Eq.~\ref{eq:x'}}


        $Z \gets \mathbb{F}(A, AGG(S', X'))$ \hfill $\triangleright$ \CommentSty{Eq.~\ref{eq:Z}}


        $Z^H \gets \mathbb{F}(A^H, AGG(S', X'))$ \hfill $\triangleright$ \CommentSty{Eq.~\ref{eq:ZH}}

        $\ell(z_i^{1}), \ell(z_i^{2})$ are obtained by calculating the similarity between positive samples and the similarity between negative samples with ($Z, Z^H$)
        

        $\mathcal{L} \gets \frac{1}{2N}\sum_{i=1}^N[\ell(z_i^{1})+\ell(z_i^{2})]$ \hfill $\triangleright$ \CommentSty{Eq.~\ref{eq:loss}}
        
        update the GCLS$^2$ network parameters by gradient descent to minimize $\mathcal{L}$

    }

    \Return $Z$
    
\end{algorithm}

\subsection{Theoretical Analyses}
In this section, we provide a theoretical justification for the GCLS$^2$ model from Mutual Information (MI) perspective.

We find that the absolute value of our proposed loss Eq.~\ref{eq:loss} is a lower bound on the mutual information maximization between the node features and the embeddings of view 1 and view 2, which has been widely used in representation learning tasks~\cite{zhu2020deep, bachman2019learning, tschannenmutual, poole2019variational}.

\begin{table*}[h]
    \caption{Statistics of the tested real-world graph datasets.}
    \label{tab: datasets}
    \begin{center}
    \resizebox{0.90\linewidth}{!}{
    \begin{tabular}{cccccccc}
        \hline
        \textbf{Datasets} & \textbf{Type} & \textbf{\#Nodes} & \textbf{\#Edges} & \textbf{Avg. Deg.} & \textbf{\#Attributes} & \textbf{\#Communities} & \textbf{\#Partitions} \\
        \hline
        Cora & Citation Network & 2,708 & 5,429 & 4.01 & 1,433 & 7 & 1\\
        Citeseer & Citation Network & 3,327 & 4,732 & 2.84 & 3,703 & 6 & 1\\
        PubMed & Citation Network & 19,717 & 44,338 & 4.50 & 500 & 3 & 1\\
        CoauthorCS & Citation Network & 18,333 & 81,894 & 8.93 & 6,805 & 15 & 1\\
        AmazonPhotos & Social Network & 7,650 & 119,081 & 31.13 & 745 & 8 & 1\\
        Email-Eu & Social Network & 1,005 & 25,571 & 50.93 & NA & 42 & 1\\
        TWeibo & Social Network & 1,944,589 & 50,133,382 & 51.56 & 1,657 & 8 & 256\\
        AmazonProducts & Social Network & 1,569,960 & 132,954,714 & 169.37 & 200 & 107 & 256\\
        \hline
    \end{tabular}
    }
    \end{center}
    
\end{table*}

\begin{lemma}
    Let $\mathbf{X}$ be a variable with random uniform distribution after GNN coding, where $\mathbf{X}_i$ is the output vector of the node within the domain indicated by the GNN architecture. Given two embedding representations $\mathbf{Z}^1,\mathbf{Z}^2 \in \mathbb{R}^{L}$ of view 1 and view 2 are two random variables, their joint distribution is denoted as $p(\mathbf{Z}^1,\mathbf{Z}^2)$. Our objective $\mathcal{L}$ is a lower bound on the mutual information of $\mathbf{X}$ and $(\mathbf{Z}^1,\mathbf{Z}^2)$ as follows, 
    \begin{equation}
        -\mathcal{L} \leq I(\mathbf{X}; \mathbf{Z}^1,\mathbf{Z}^2).
    \end{equation}
    \label{lemma:tA}
\end{lemma}

\begin{proof}
    For a rigorous comparison, we compare our objective with the GRACE objective~\cite{oord2018representation, poole2019variational, zhu2020deep}, which is defined under existing conditions as follows, 
    \begin{equation}
        \Gamma(z^1_i)=\sum_{j=1}^{N}e^{sim(z_i^1,z_j^2)/\tau},
    \end{equation}
    \begin{equation}
        \gamma(z^1_i)=\sum_{j=1, j \neq i}^{N}e^{sim(z_i^1,z_j^1)/\tau},
    \end{equation}
    \begin{equation}
        \label{eq:ACE}
        \small
        I_{ACE}(\mathbf{Z}^1;\mathbf{Z}^2) \triangleq \mathbb{E}_{\prod_{i}p(z_i^1,z_i^2) }[\frac{1}{N} \sum_{i=1}^{N} log \frac{e^{sim(z_i^1,z_i^2)/\tau}}{\Gamma(z^1_i)+\gamma(z^1_i)}],
    \end{equation}
    where $\prod_{i}p(z_i^1,z_i^2)$ denotes the joint distribution. To facilitate with the notational representation, we define $\Gamma(z^1_i)=\sum_{j=1}^{N}e^{sim(z_i^1,z_j^2)/\tau}$, $\gamma(z^1_i)=\sum_{j=1, j \neq i}^{N}e^{sim(z_i^1,z_j^1)/\tau}$. Please note that $\Gamma(z^2_i)$ and $\gamma(z^2_i)$ are defined symmetrically with $\Gamma(z^1_i)$ and $\gamma(z^1_i)$ respectively. 
    So that our objective $\mathcal{L}$ can be rewritten as
    \begin{equation}
        \small
        \label{eq:L}
        -\mathcal{L} = \mathbb{E}_{\prod_{i}p(z_i^1,z_i^2) }[\frac{1}{N} \sum_{i=1}^{N} log \frac{pos(z_i^1)+e^{sim(z_i^1,z_i^2)/\tau}}{\sqrt{(\Gamma(z^1_i)+\gamma(z^1_i))(\Gamma(z^2_i)+\gamma(z^2_i))}}],
    \end{equation}
    where the $\sqrt{(\Gamma(z^1_i)+\gamma(z^1_i))(\Gamma(z^2_i)+\gamma(z^2_i))}$ is to illustrate the symmetry of $\Gamma(z^1_i)+\gamma(z^1_i)$ and $\Gamma(z^2_i)+\gamma(z^2_i)$, which are equal in value. Therefore, we can use the Eq.~\ref{eq:ACE} Eq.~\ref{eq:L} to deduce that
    \begin{equation}
    \label{LACE}
    \small
    \begin{split}
        -2\mathcal{L} &= I_{ACE}(\mathbf{Z}^1;\mathbf{Z}^2) - \mathbb{E}_{\prod_{i}p(z_i^1,z_i^2) }[ \frac{1}{N} \sum_{i=1}^{N} log (1 + \frac{pos(z_i^1)}{e^{sim(z_i^1,z_i^2)/\tau}})] \\ &+ I_{ACE}(\mathbf{Z}^2;\mathbf{Z}^1) - \mathbb{E}_{\prod_{i}p(z_i^1,z_i^2) }[ \frac{1}{N} \sum_{i=1}^{N} log (1 + \frac{pos(z_i^2)}{e^{sim(z_i^2,z_i^1)/\tau}})] \\
        &\leq  I_{ACE}(\mathbf{Z}^1;\mathbf{Z}^2) + I_{ACE}(\mathbf{Z}^2;\mathbf{Z}^1).
    \end{split}
    \end{equation}
    And according to~\cite{zhu2020deep, poole2019variational}, the GRACE objective is a lower bound of the true MI, i.e.,
    \begin{equation}
        I_{ACE}(\mathbf{Z}^1;\mathbf{Z}^2) \leq I(\mathbf{Z}^1;\mathbf{Z}^2).
    \end{equation}
    Thus, we have the following inference
    \begin{equation}
        -2\mathcal{L} \leq  I(\mathbf{Z}^1;\mathbf{Z}^2) + \leq I(\mathbf{Z}^2;\mathbf{Z}^1) \leq  2I(\mathbf{Z}^1;\mathbf{Z}^2).
    \end{equation}
    This is equivalent to
    \begin{equation}
        -\mathcal{L} \leq I(\mathbf{Z}^1;\mathbf{Z}^2).
        \label{eq:12}
    \end{equation}
    For three random variables $\mathbf{X}$, $\mathbf{Y}$, $\mathbf{Z}$ satisfying one Markov chain $\mathbf{X} \rightarrow \mathbf{Y} \rightarrow \mathbf{Z}$, we have $I(\mathbf{X};\mathbf{Z}) \leq I(\mathbf{X};\mathbf{Y})$ holds.
    We find that $\mathbf{X}$, $\mathbf{Z}^1$, $\mathbf{Z}^2$ satisfy the chain relation $\mathbf{Z}^1 \leftarrow \mathbf{X} \rightarrow \mathbf{Z}^2$.
    Actuality, $\mathbf{Z}^1$ and $\mathbf{Z}^2$ are conditionally independent after considering $\mathbf{X}$. This represents the existence of Markov chain $\mathbf{Z}^1 \rightarrow \mathbf{X} \rightarrow \mathbf{Z}^2$, and we have $I(\mathbf{Z}^1;\mathbf{Z}^2) \leq I(\mathbf{Z}^1;\mathbf{X})$.
    Since the chain relation $\mathbf{X} \rightarrow (\mathbf{Z}^1,\mathbf{Z}^2) \rightarrow \mathbf{Z}^1$ holds, it follows that $I(\mathbf{X};\mathbf{Z}^1) \leq I(\mathbf{X};\mathbf{Z}^1,\mathbf{Z}^2)$.
    Combining the two inequalities above, we have the inequality
    \begin{equation}
        I(\mathbf{Z}^1;\mathbf{Z}^2) \leq I(\mathbf{X};\mathbf{Z}^1,\mathbf{Z}^2).
        \label{eq:123}
    \end{equation}
    According to Eq.~\ref{eq:12} and Eq.~\ref{eq:123}, we can derive the final inequality as follows and complete the proof,
    \begin{equation}
        -\mathcal{L} \leq I(\mathbf{X}; \mathbf{Z}^1,\mathbf{Z}^2).
    \end{equation}
\end{proof}

\noindent
\textbf{Analysis: } Through Lemma~\ref{lemma:tA}, we can know that minimizing $\mathcal{L}$ is equivalent to maximizing the lower bound on the mutual information $I(\mathbf{X};\mathbf{Z}^1,\mathbf{Z}^2)$ between the input node features and the learned node representations, which theoretically provides the success of our model training. 
This is what the previous work~\cite{gong2023ma, hu2020gpt, shen2023neighbor, chen2023attribute, xin2017deep} on contrastive learning has not focused on.

\begin{table*}[!t]
    \centering
    \caption{Comparison between different baselines and GCLS$^2$ on different graph networks with attributes. The best performance is \textbf{highlighted} in bold, and the second best is \underline{underlined}.}
    \label{performance}
    \resizebox{\linewidth}{!}{
    \begin{tabular}{cc|cccccccc|c}
        \toprule
        \multirow{2}{1cm}[-0.5ex]{\bf Datasets} & \multirow{2}{*}[-0.5ex]{\bf Metrics} & \multicolumn{9}{c}{\bf Methods}\\
        \cmidrule{3-11}
        & & GCN  & DGI & MVGRL & GRACE & GCA & SUGRL & NCLA & ASP & {\bf GCLS$^2$} \\
        \midrule
        \multirow{3}{*}{\bf Cora} 
        & ACC & 81.53$\pm$0.88 & 82.87$\pm$0.14 & 83.24$\pm$1.55 & 79.48$\pm$3.54 & 81.93$\pm$1.43 & 83.35$\pm$1.24 & 79.77$\pm$2.51 & \underline{84.75$\pm$0.84} & \textbf{87.82$\pm$0.93} \\
        & NMI & 66.33$\pm$1.33 & 63.41$\pm$0.25 & 67.89$\pm$0.79 & 58.87$\pm$4.77 & 67.98$\pm$1.13 & 70.67$\pm$0.57 & 59.30$\pm$5.95 & \underline{71.54$\pm$0.64} & \textbf{74.11$\pm$1.62} \\
        & MF1  & 81.64$\pm$0.74 & 82.71$\pm$0.15 & 83.36$\pm$1.68 & 79.41$\pm$3.56 & 81.57$\pm$1.59 & 82.78$\pm$1.54 & 79.71$\pm$2.38 & \underline{84.34$\pm$1.44} & \textbf{87.90$\pm$0.96} \\
        \midrule
        \multirow{3}{*}{\bf Citeseer} 
        & ACC & 71.34$\pm$1.47 & 71.83$\pm$0.64 & 72.36$\pm$1.35 & 71.38$\pm$1.59 & 70.67$\pm$2.45 & 72.08$\pm$1.84 & 71.48$\pm$1.29 & \textbf{72.69$\pm$1.14} & \underline{72.64$\pm$1.22} \\
        & NMI & 46.13$\pm$0.77 & 45.79$\pm$0.48 & 46.34$\pm$0.56 & 45.23$\pm$3.01 & 45.97$\pm$1.08 & 44.35$\pm$0.67 & 46.29$\pm$3.57 & \underline{46.64$\pm$0.54} & \textbf{46.88$\pm$0.84} \\
        & MF1  & 72.08$\pm$0.85 & 71.97$\pm$0.41 & 72.48$\pm$1.64 & 71.88$\pm$1.64 & 71.14$\pm$2.76 & 72.28$\pm$0.44 & \underline{72.89$\pm$1.15} & 72.34$\pm$1.24 & \textbf{73.05$\pm$0.70} \\
        \midrule
        \multirow{3}{*}{\bf PubMed} 
        & ACC & 80.39$\pm$1.05 & 76.88$\pm$0.65 & 79.81$\pm$1.03 & 80.74$\pm$1.20 & 81.77$\pm$2.50 & \underline{82.57$\pm$1.64} & 80.55$\pm$1.44 & 80.74$\pm$0.64  & \textbf{83.22$\pm$0.34} \\
        & NMI & 41.36$\pm$2.15 & 37.68$\pm$0.82 & 42.07$\pm$1.52 & 42.08$\pm$2.30 & 41.26$\pm$2.10 & 44.76$\pm$0.84 & 41.74$\pm$2.82 & \underline{45.33$\pm$0.54}  & \textbf{47.29$\pm$0.75} \\
        & MF1  & 80.38$\pm$1.06 & 76.34$\pm$1.15 & 80.13$\pm$0.96 & 80.52$\pm$1.25 & 82.34$\pm$0.76 & \textbf{83.48$\pm$0.90} & 80.54$\pm$0.47 & 81.34$\pm$1.34  & \underline{83.17$\pm$0.34} \\
        \midrule
        \multirow{3}{*}{\bf \shortstack{Coauthor\\-CS}} 
        & ACC & 82.59$\pm$0.87 & \textbf{92.03$\pm$0.54} & \underline{91.56$\pm$0.52} & 84.25$\pm$0.58 & 90.91$\pm$1.12 & 91.23$\pm$0.91 & 85.35$\pm$0.41 & 87.64$\pm$0.52  & 90.08$\pm$0.20 \\
        & NMI & 68.52$\pm$1.31 & \textbf{81.98$\pm$1.35} & 80.60$\pm$1.42 & 71.10$\pm$0.77 & 81.03$\pm$0.91 & \underline{81.48$\pm$0.34} & 72.79$\pm$0.98 & 79.13$\pm$0.74  & 80.77$\pm$0.97 \\
        & MF1  & 82.43$\pm$0.86 & \textbf{91.78$\pm$0.93} & \underline{91.60$\pm$0.72} & 84.18$\pm$0.71 & 90.62$\pm$1.58 & 91.48$\pm$0.94 & 87.30$\pm$0.66 & 71.34$\pm$0.24  & 90.11$\pm$0.57 \\
        \midrule
        \multirow{3}{*}{\bf \shortstack{Amazon\\-Photo}} 
        & ACC & 82.41$\pm$1.51 & 85.48$\pm$1.23 & 87.61$\pm$1.37 & 83.39$\pm$1.57 & 88.02$\pm$1.93 & \underline{89.42$\pm$0.71} & 83.15$\pm$1.29 & 86.94$\pm$0.76  & \textbf{89.55$\pm$1.16} \\
        & NMI & 64.69$\pm$3.98 & 74.48$\pm$0.47 & 76.73$\pm$0.82 & 66.75$\pm$2.03 & 76.92$\pm$1.85 & \underline{77.78$\pm$0.45} & 66.30$\pm$2.01 & 75.28$\pm$0.33 & \textbf{77.88$\pm$1.81} \\
        & MF1 & 82.45$\pm$1.43 & 85.59$\pm$1.44 & 87.43$\pm$1.64 & 83.32$\pm$1.70 & 88.73$\pm$2.32 & \underline{89.68$\pm$0.37} & 83.06$\pm$1.35 & 87.14$\pm$1.21 &  \textbf{89.87$\pm$1.11} \\
        \bottomrule
    \end{tabular}
    }
    
\end{table*}

\begin{table*}[!t]
    \centering
    \caption{Comparison between different baselines and GCLS$^2$ on different graph networks without attributes. The best performance is \textbf{highlighted} in bold, and the second-best is \underline{underlined}.}
    \label{performance-withoutF}
    \resizebox{\linewidth}{!}{
    \begin{tabular}{cc|cccccccc|c}
        \toprule
        \multirow{2}{1cm}[-0.5ex]{\bf Datasets} & \multirow{2}{*}[-0.5ex]{\bf Metrics} & \multicolumn{9}{c}{\bf Methods}\\
        \cmidrule{3-11}
        & & GCN & DGI & MVGRL & GRACE & GCA & SUGRL & NCLA & ASP & {\bf GCLS$^2$} \\
        \midrule
        \multirow{3}{*}{\bf Cora} 
        & ACC & 76.75$\pm$4.43 & 79.18$\pm$0.57 & 80.63$\pm$1.78 & 78.54$\pm$0.86 & 78.55$\pm$1.53 & 82.58$\pm$0.67 & 77.92$\pm$1.64 & \underline{83.25$\pm$1.49} & \textbf{85.05$\pm$0.83} \\
        & NMI & 55.02$\pm$5.09 & 68.14$\pm$0.94 & 66.50$\pm$1.22 & 67.34$\pm$0.79 & 52.56$\pm$1.65 & 71.43$\pm$0.83 & 58.91$\pm$1.41 & \underline{71.68$\pm$0.78}  & \textbf{73.94$\pm$1.62} \\
        & MF1 & 76.35$\pm$4.64 & 79.48$\pm$0.74 & 79.80$\pm$1.62 & 78.34$\pm$0.96 & 77.92$\pm$1.46 & 81.68$\pm$0.74 & 76.60$\pm$1.39 & \underline{84.34$\pm$2.24} &  \textbf{87.71$\pm$0.96} \\
        \midrule
        \multirow{3}{*}{\bf Citeseer} 
        & ACC & 58.40$\pm$2.86 & 61.35$\pm$0.48 & 62.60$\pm$1.12 & 60.38$\pm$1.59 & 61.96$\pm$1.34 & \textbf{63.48$\pm$0.74} & 61.48$\pm$1.29 & \underline{63.34$\pm$0.43} &  62.40$\pm$1.26 \\
        & NMI & 26.95$\pm$2.96 & 32.48$\pm$0.94 & 34.60$\pm$1.02 & 32.83$\pm$2.08 & 31.49$\pm$0.90 & \textbf{36.58$\pm$0.49} & 33.29$\pm$3.01 & 35.76$\pm$0.59 &  \underline{35.77$\pm$1.10} \\
        & MF1  & 58.27$\pm$3.11 & 61.48$\pm$0.46 & 62.80$\pm$1.52 & 60.88$\pm$1.64 & 61.28$\pm$1.15 & \underline{64.18$\pm$0.53} & 61.89$\pm$1.18 & \textbf{64.34$\pm$0.72} &  62.12$\pm$1.13 \\
        \midrule
        \multirow{3}{*}{\bf PubMed} 
        & ACC & 77.73$\pm$0.87 & 75.68$\pm$0.74 & 78.40$\pm$1.57 & 76.38$\pm$0.46 & 79.81$\pm$1.01 & 80.48$\pm$0.94 & 76.87$\pm$1.63 & \underline{81.74$\pm$0.64} &  \textbf{82.31$\pm$0.52} \\
        & NMI & 36.68$\pm$1.64 & 35.24$\pm$0.46 & 38.60$\pm$1.28 & 37.34$\pm$0.84 & 42.27$\pm$1.36 & 41.65$\pm$0.82 & 37.60$\pm$1.32 & \underline{43.34$\pm$0.58}  & \textbf{45.38$\pm$1.12} \\
        & MF1  & 77.60$\pm$1.85 & 75.48$\pm$0.47 & 78.64$\pm$1.73 & 76.74$\pm$0.29 & 79.29$\pm$0.93 & 80.95$\pm$0.36 & 76.70$\pm$1.02 & \underline{81.34$\pm$0.85}  & \textbf{82.33$\pm$0.54} \\
        \midrule
        \multirow{3}{*}{\bf \shortstack{Coauthor\\-CS}} 
        & ACC & 80.87$\pm$0.98 & 87.78$\pm$0.36 & 87.62$\pm$1.32 & 84.09$\pm$0.74 & \underline{87.89$\pm$0.82} & 87.48$\pm$0.63 & 84.53$\pm$0.63 & 86.34$\pm$0.84 & \textbf{89.30$\pm$0.62} \\
        & NMI & 66.67$\pm$1.69 & \underline{77.45$\pm$0.46} & 76.58$\pm$1.53 & 70.74$\pm$1.16 & 75.68$\pm$1.57 & 75.18$\pm$0.65 & 71.68$\pm$0.72 & 76.33$\pm$0.76 & \textbf{79.35$\pm$0.83} \\
        & MF1 & 80.74$\pm$0.88 & 87.46$\pm$0.62 & 87.40$\pm$1.25 & 84.05$\pm$0.70 & 87.28$\pm$1.33 & \underline{87.82$\pm$0.65} & 84.50$\pm$0.97 & 86.34$\pm$0.98 & \textbf{89.37$\pm$0.62} \\
        \midrule
        \multirow{3}{*}{\bf \shortstack{Amazon\\-Photo}} 
        & ACC & 80.71$\pm$1.21 & 85.47$\pm$0.72 & 86.69$\pm$1.53 & 85.49$\pm$0.87 & 86.70$\pm$1.45 & \underline{87.58$\pm$0.52} & 85.45$\pm$1.04 & 83.76$\pm$0.85 &  \textbf{88.98$\pm$0.56} \\
        & NMI & 65.31$\pm$1.62 & 71.48$\pm$0.44 & 73.50$\pm$1.66 & 70.57$\pm$1.88 & 75.05$\pm$1.22 & \underline{75.42$\pm$0.63} & 70.61$\pm$1.58 & 74.34$\pm$0.75 &  \textbf{75.91$\pm$1.69} \\
        & MF1 & 80.65$\pm$0.96 & 85.48$\pm$0.63 & 86.50$\pm$1.42 & 85.44$\pm$0.79 & 87.01$\pm$2.10 & \underline{87.47$\pm$0.83} & 85.35$\pm$1.35 & 83.44$\pm$0.55 &  \textbf{88.89$\pm$0.67} \\
        \midrule
        \multirow{3}{*}{\bf \shortstack{Email\\-Eu}} 
        & ACC & 62.90$\pm$2.30 & 64.33$\pm$0.87 & 65.45$\pm$1.46 & 63.79$\pm$3.79 & 67.30$\pm$2.69 & 64.37$\pm$0.93 & 64.32$\pm$1.52 & \underline{68.52$\pm$2.73} &  \textbf{68.90$\pm$4.10} \\
        & NMI & 79.93$\pm$1.58 & 78.24$\pm$0.56 & 80.61$\pm$1.31 & 78.83$\pm$0.45 & \underline{81.93$\pm$2.13} & 79.48$\pm$0.67 & 80.60$\pm$1.64 & 81.34$\pm$0.98 &  \textbf{82.41$\pm$1.37} \\
        & MF1  & 62.41$\pm$2.43 & 65.18$\pm$0.64 & 65.80$\pm$1.32 & 63.34$\pm$3.24 & 67.28$\pm$2.10 & 64.48$\pm$0.84 & 65.60$\pm$1.62 & \underline{69.33$\pm$2.53} &  \textbf{72.02$\pm$3.42} \\
        \bottomrule
    \end{tabular}
    }
    
\end{table*}

\section{Experimental Evaluation}
\label{sec:EX}
In this section, we report the effectiveness and efficiency of our proposed GCLS$^2$ approach on real-world graph datasets by answering the following four research questions (RQs).

\begin{itemize}
\item {\bf RQ1 (Superiority of GCLS$^2$)}: What are the advantages of GCLS$^2$ compared with state-of-the-art methods? 

\item {\bf RQ2 (Applicability of GCLS$^2$)}: What are impacts of different parameter settings (e.g., semantic dimension $d$ and temperature parameter $\tau$) on the GCLS$^2$ performance? 

\item {\bf RQ3 (Scalability of GCLS$^2$)}: Can GCLS$^2$+HGP algorithm perform the community detection efficiently on large-scale graphs? 

\item {\bf RQ4 (Benefits of GCLS$^2$)}:  Can GCLS$^2$ significantly improve the accuracy, efficiency, and modularity of the community detection?

\end{itemize}

\subsection{Experimental Settings}

\noindent \textbf{Datasets and Baselines: }
We use six public real-world graph datasets: 
the widely-used citation networks (i.e., Cora, Citeseer, and PubMed~\cite{sen2008collective}),
the co-authorship network and the product co-purchase network (i.e., CoauthorCS and AmazonPhotos~\cite{shchur2018pitfalls}, resp.), 
and the e-mail network between members of a European research institution (Email\_Eu~\cite{YinBLG17}) 
to evaluate the effectiveness of our GCLS$^2$ method on the community detection task.
We also use two large-scale public real-world graphs, TWeibo~\cite{yang2023pane} (i.e., a social network of Tencent Weibo users) and AmazonProducts~\cite{GraphSAINT} (i.e., a product review network of Amazon), to demonstrate the applicability and effectiveness of our proposed high-level graph partitioning algorithms HGP and GCLS$^2$.
The details of the datasets are in Table~\ref{tab: datasets}.
We use the notation ``NA'' to denote the absence of attributes for each node in the graph dataset.
We compare our approach with two supervised learning baselines (i.e., GCN~\cite{kipf2016semi} and GAT~\cite{gat}) and seven unsupervised learning baselines, including DGI~\cite{velivckovicdeep}, MVGRL~\cite{hassani2020contrastive}, GRACE~\cite{zhu2020deep}, GCA~\cite{zhu2021graph}, SUGRL~\cite{mo2022simple}, NCLA~\cite{shen2023neighbor}, and ASP~\cite{chen2023attribute}. 
We compare the HGP in Algorithm~\ref{alg:hgp} with three classical graph partitioning algorithms (i.e., METIS~\cite{karypis1998fast}, LGD~\cite{stanton2012streaming}, and Hash~\cite{abbas2018streaming} algorithms) on large-scale graphs, in terms of the efficiency of graph partitioning and the accuracy of the community detection.

\noindent \textbf{Metrics:}
We use classical community detection metrics, \textit{normalized mutual information} (NMI), \textit{accuracy} (ACC), and \textit{Macro-F1 score} (MF1)~\cite{kong2011robust}, to evaluate the effectiveness of the model detection. For all the above metrics, the higher their values, the better. 
To evaluate the efficiency of GCLS$^2$ and HGP, we measured the \textit{wall clock times} of online community detection and graph partitioning, respectively. We evaluate the load balancing capability of graph partitioning algorithms by calculating the \textit{maximum load difference} between partitions; the smaller the \textit{maximum load difference}, the better the load balancing capability.

\begin{table}[!t]
    \centering
    \caption{Performance of GCLS$^2$ on large-scale graph datasets.}
    \label{tab:large-d}
    \resizebox{\linewidth}{!}{
    \begin{tabular}{cccccccc}
        \toprule
        \multirow{3}{*}[-0.5ex]{\bf Methods} & \multicolumn{3}{c}{\bf TWeibo} & \phantom{a}
        & \multicolumn{3}{c}{\bf AmazonProducts}\\
        \cmidrule{2-4} \cmidrule{6-8}
        & ACC &  NMI  & MF1 & & ACC & NMI & MF1\\
        \midrule
        {\bf \shortstack{GCN}} & 33.34 & 20.56 & 38.39 & & 30.52 & 23.75 & 38.12\\
        {\bf \shortstack{GAT}} & 35.78 & 24.26 & 36.91 & & 41.72 & 36.87 & 43.57\\
        {\bf \shortstack{MVGRL}} & 38.68 & 30.47 & 37.77 & & 50.32 & 43.48 & 51.46\\
        {\bf \shortstack{GRACE}} & 35.20 & 28.92 & 26.19 & & 46.05 & 40.94 & 49.31\\
        {\bf \shortstack{SUGRL}} & \underline{39.57} & \underline{34.25} & \underline{45.79} & & \underline{54.93} & \underline{47.82} & \underline{55.68}\\
        \midrule
        {\bf \shortstack{GCLS$^2$}} & \textbf{42.94} & \textbf{44.72} & \textbf{59.92} & & \textbf{61.79} & \textbf{53.01} & \textbf{63.54}\\
        \bottomrule
    \end{tabular}
    }
\end{table}

\begin{figure}[!t]
    \centering
    \includegraphics[width=0.95\linewidth]{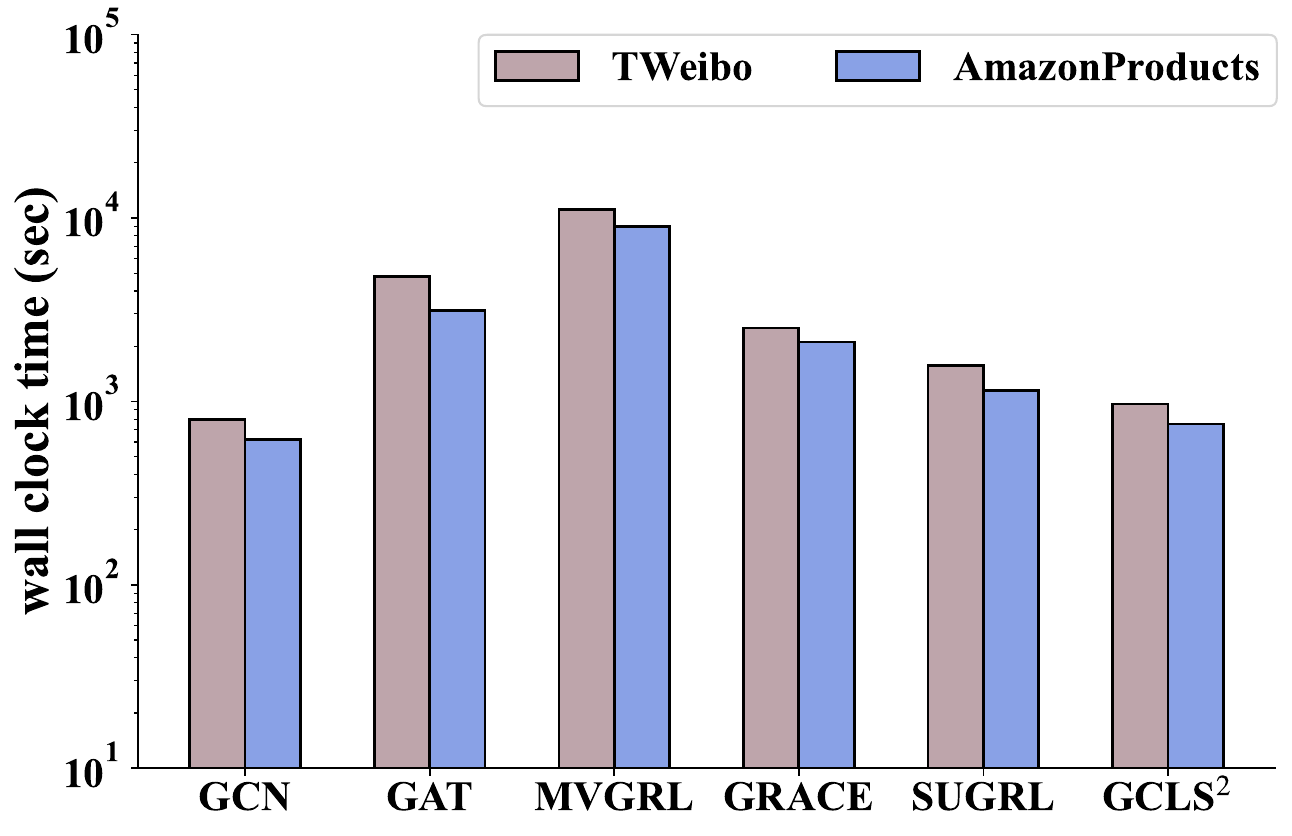}
    \caption{The community detection inference time comparison of different methods on large-scale graph datasets.}
    \label{fig:runtime}
\end{figure}

\noindent \textbf{Default Parameters:}
For every experiment, each model is first trained in an unsupervised manner. 
During the graph contrastive learning phase, we employ the SSS encoder with 500 epochs, using the Adam optimizer and a learning rate of 5$e$-4.
We select the hyperparameters with the best experimental results as the default hyperparameters. 
Specifically, the temperature parameter $\tau$ is set to 1, the structure semantic encoder dimension $d$ is set to 32, and the partition size $p$ is set to 256.
In the community detection phase, we employ a 2-layer GCN for training, with 500 epochs. The nodes of each dataset are divided into a training set, a validation set, and a testing set in a proportion of 8: 1: 1, and all models are set with an early stop strategy.

We run all the experiments on a PC with an NVIDIA RTX 4090 GPU and 24GB of memory. All algorithms are implemented in Python and executed with the Python 3.11 interpreter. 


\begin{table}[!t]
    \centering
    \caption{Performance of the variants on Cora and Email-Eu.}
    \label{tab:ablation}
    \resizebox{\linewidth}{!}{
    \begin{tabular}{cccccccc}
        \toprule
        \multirow{2}{*}[-0.5ex]{\bf Variants} & \multicolumn{3}{c}{\bf Cora} & \phantom{a}
        & \multicolumn{3}{c}{\bf Email-Eu}\\
        \cmidrule{2-4} \cmidrule{6-8}
        & ACC &  NMI  & MF1 & & ACC & NMI & MF1\\
        \midrule
        {\bf \shortstack{GCLS$^2$ w/o $\boldsymbol S$}} & 86.23 & 72.08 & 86.30 & & \underline{65.50} & 80.23 & \underline{66.73}\\
        {\bf \shortstack{GCLS$^2$ w/o SSS}} & \underline{86.67} & \underline{72.84} & 86.74 & & 64.70 & \underline{80.65} & 64.61\\
        {\bf \shortstack{GCLS$^2$ w/o SCL}} & 86.63 & 72.72 & \underline{86.75} & & 59.99 & 79.73 & 59.62\\
        \midrule
        {\bf \shortstack{GCLS$^2$}} & \textbf{87.82} & \textbf{74.11} & \textbf{87.90} & & \textbf{68.90} & \textbf{82.41} & \textbf{72.02}\\
        \bottomrule
    \end{tabular}
    }
    
\end{table}

\begin{figure*}
    \centering
    \subfigure[partitioning efficiency]{\label{fig:hgp-t}\includegraphics[width=0.30\linewidth]{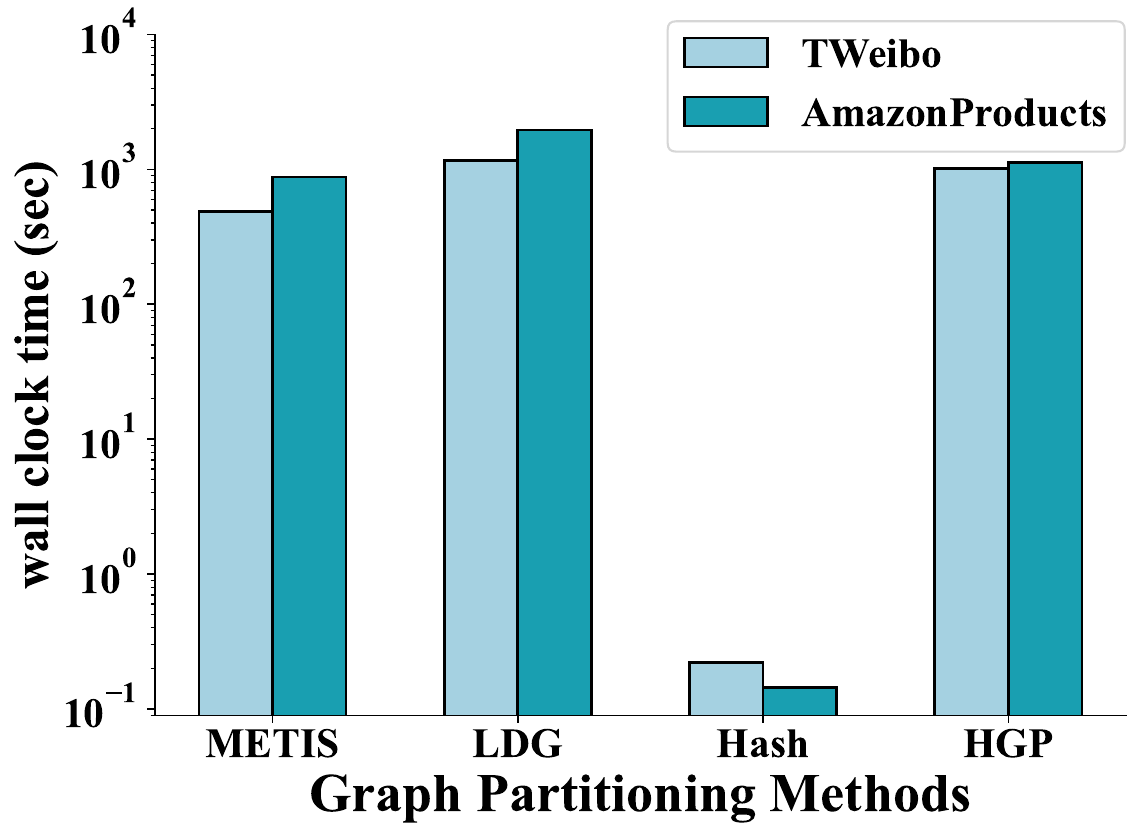}}\hspace{0.5cm}
    \subfigure[load difference]{\label{fig:hgp-l}\includegraphics[width=0.30\linewidth]{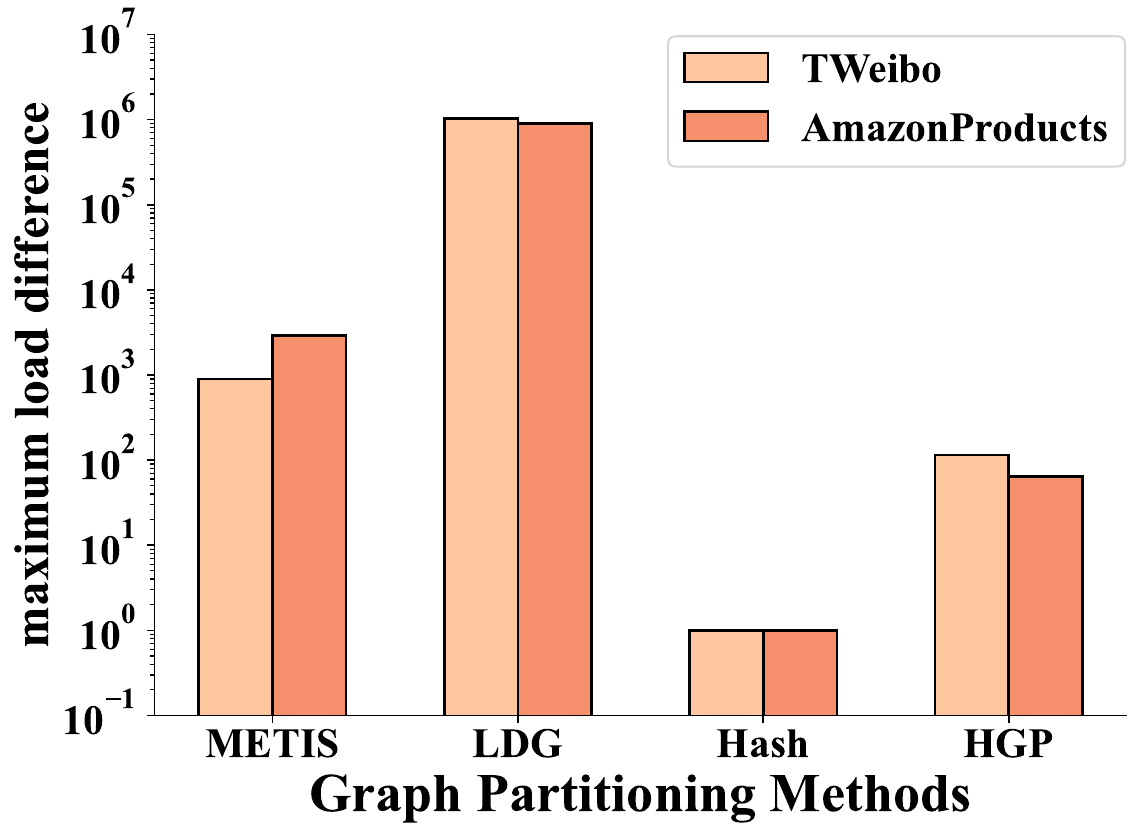}}\hspace{0.5cm}
    \subfigure[analysis of p-parameters]{\label{fig:hgp-p}\includegraphics[width=0.30\linewidth]{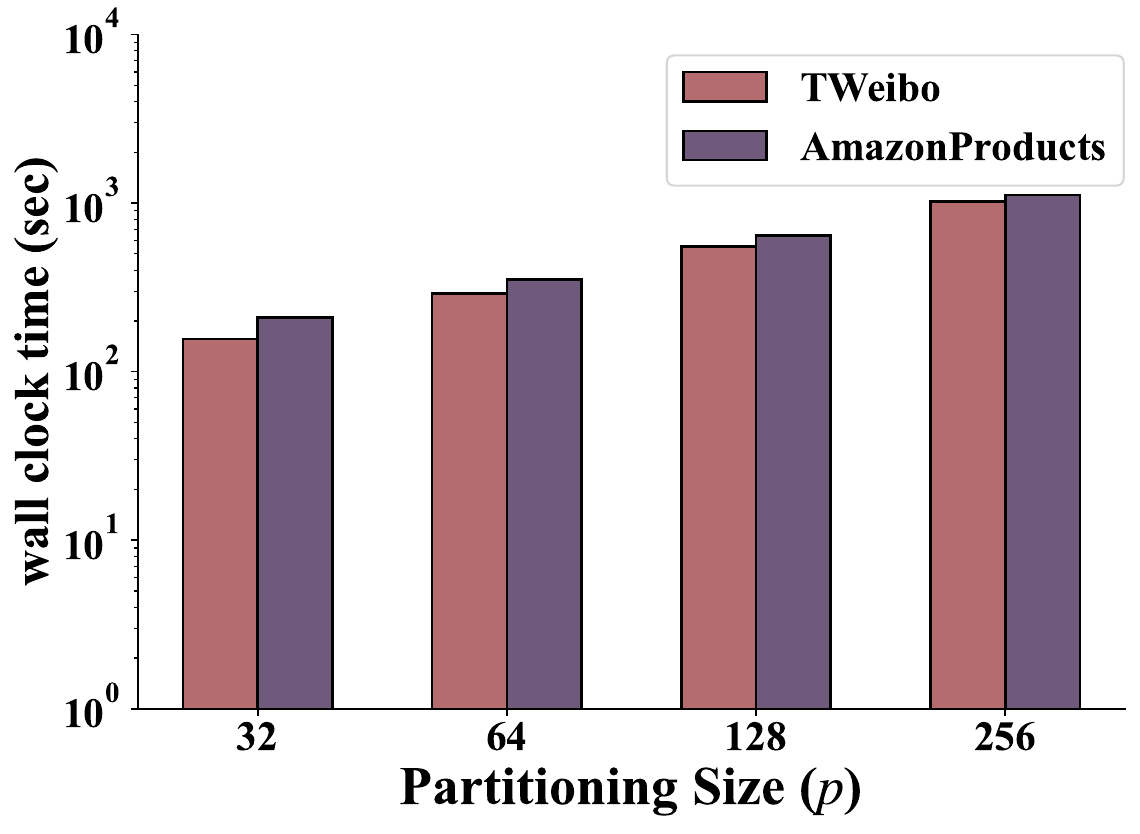}}
    \caption{Performance evaluation of the HGP algorithm.}
    \label{fig:hdg}
\end{figure*}

\subsection{The Performance Evaluation -- RQ1}
To answer RQ1, we evaluate the performance of the baselines and GCLS$^2$ with and without attributes. For the without-attribute case, we test by replacing the attribute matrix with an adjacency matrix.
Table~\ref{performance} shows the performance of the baselines and GCLS$^2$ on different graph networks with attributes.
Table~\ref{performance-withoutF} shows the performance without attributes.
Overall, the tables show that our GCLS$^2$ performs strongly on all six datasets, especially in the case of without attributes, where it has quite an advantage. These performances validate the excellence of our GCLS$^2$ method. Specifically, we have the following observations.

First, GCLS$^2$ has an advantage in comparison to GCN, which we believe is due to the fact that GCLS$^2$ deeply mines the structure information of the community and enhances the representation of this information through structure contrastive learning, and also compared to GCN, which has an average reduction of 4.76\% in the case of without attributes, GCLS$^2$ has a reduction of 3.05\%, where excluding the Citeseer dataset, it has only a reduction of 1\%. This emphasizes the importance of considering structure semantics in community detection.

Second, we find that the node-graph level contrastive learning methods (i.e., DGI, MVGRL) are significantly more effective than the node-node level (i.e., GRACE, NCLA, GCLS$^2$) in the CoauthorCS dataset. 
We believe that the CoauthorCS with a large number of attributes (i.e., 6,805) is represented on the data graph with rich information, which is more sensitive to the community information, and at the same time, removing the attribute information, the node-graph level effect is not outstanding, which confirms our thoughts. 
In contrast, our proposed GCLS$^2$, which also focuses on structure semantic representation at the node-node level, has the highest detection performance without attribute information.

Due to the memory limitations of the experimental machines, most network models, with their complex structures, require significant memory resources during online training. We successfully evaluate the performance of GCN, GAT, MVGRL, GRACE, and SUGRL with our proposed HGP partitioning algorithm with partition size $p=256$ on two large-scale datasets: TWeibo and AmazonProducts.
As shown in Table~\ref{tab:large-d}, our GCLS$^2$ method achieves the best performance in terms of accuracy and modularity compared to baseline methods.
On the TWeibo dataset, GCLS$^2$ outperforms the second-best method by 3\% in accuracy and 10\% in modularity.
On the dense AmazonProducts dataset, it exceeds the second-best by 7\% in accuracy, highlighting GCLS$^2$'s advantage in handling dense graphs.
Moreover, GCLS$^2$ not only outperforms supervised learning methods such as GCN and GAT, but also learns similar embeddings with potential community structures by high-level structural contrastive learning.
This approach overcomes the limitation of traditional methods, which focus on aligning embeddings from different views of the same vertex, and proves effective for community detection on dense and large-scale datasets.

To further validate the computational cost advantage of our GCLS$^2$ model, we measure the inference time of the model on different large-scale datasets.
Figure~\ref{fig:runtime} shows the inference time of our model and baselines on two large-scale datasets (i.e., TWeibo and AmazonProducts).
The results show that GCLS$^2$ achieves the shortest inference time among all GCL models. This is primarily because GCLS$^2$ avoids additional data augmentation or node feature resampling.
For GAT, the complex attention mechanism increases the computational cost of the model.
Most GCL models are designed based on GCNs, making a single GCN the fastest in terms of inference.
In comparison, GRACE generates multiple views by removing edges and masking node features; MVGRL not only generates multiple views but also uses a discriminator to measure agreement across different views; and SUGRL further samples node features. 
These operations increase inference costs. In contrast, GCLS2 uses lightweight SSS feature extraction and structure loss calculation, with no additional overhead, resulting in faster inference.

\subsection{The Ablation Study -- RQ2}
To answer RQ2, we study the variants of the proposed GCLS$^2$ method and analyze the performance of each of the important modules of Figure~\ref{fig:framework}. Specifically, we have three variants: \romannumeral1) we separate the structure similarity matrix $S$, which represents dense community information, as a variant, i.e., GCLS$^2$ w/o $S$; \romannumeral2) we separate the structure similarity semantic (SSS) module, which extracts low-level semantic representations as a variant, i.e., GCLS$^2$ w/o SSS; \romannumeral3) we separate the structure contrastive loss that reinforces the expression of community structure information as a variant, i.e., GCLS$^2$ w/o SCL, and the rest of the processing for each variant was the same as for GCLS$^2$. Table~\ref{tab:ablation} shows the performance of the GCLS$^2$'s variants on Cora and Email-Eu datasets. 
From Table~\ref{tab:ablation}, the GCLS$^2$ consistently has the highest accuracy among all the variants on the two datasets. 
Compared to the three variants, the GCLS$^2$ achieves significant gains on Email-Eu and Cora. 
Especially in the Email-Eu, the accuracy of using SCL increased by 8\%.
This reflects that the structure semantic information is effective in community detection, and the structure contrastive learning has shown good performance improvement on Email-Eu.

\begin{figure*}[!ht]
    \centering
    \subfigure{\includegraphics[width=0.28\linewidth, height=0.18\hsize]{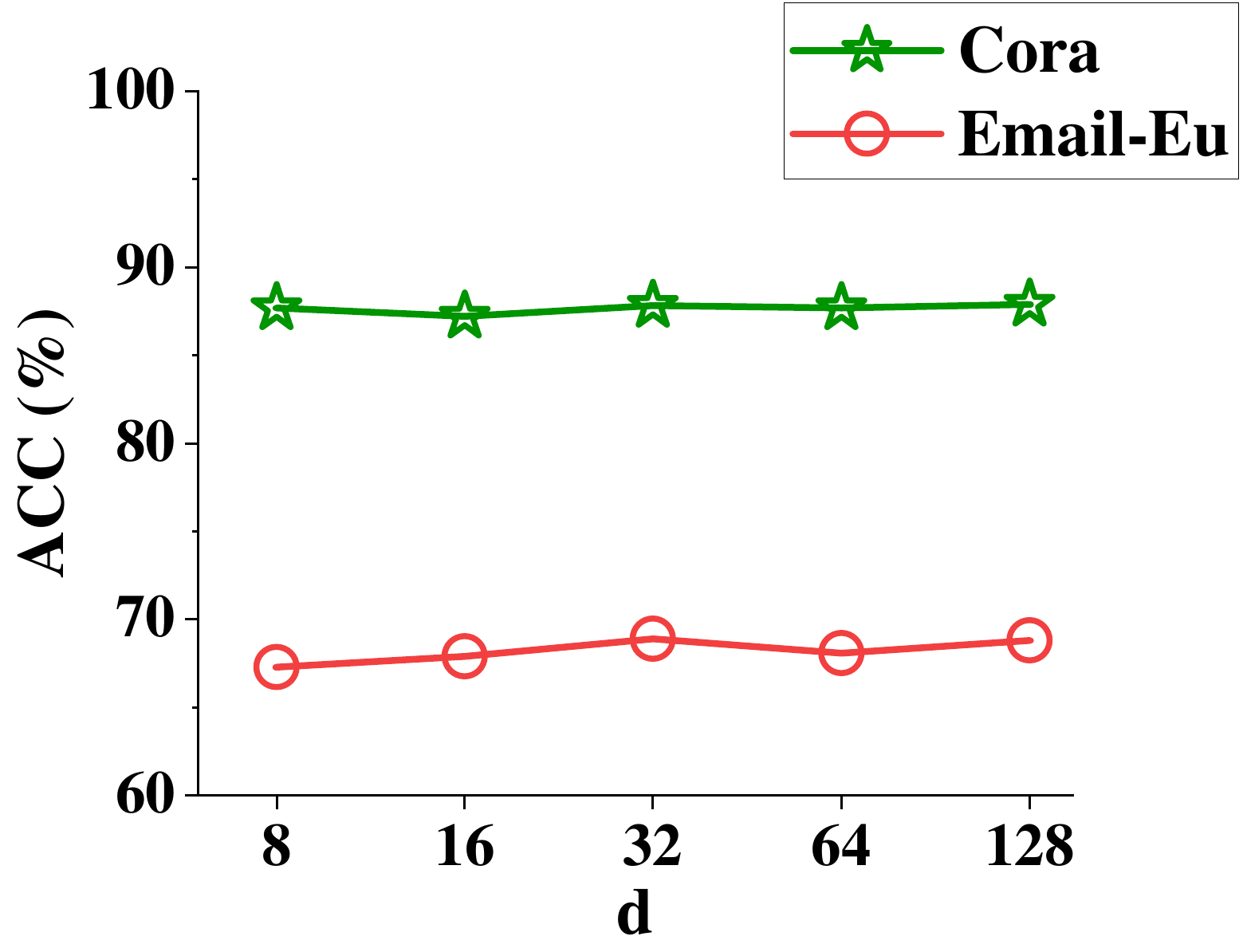}}\hspace{0.5cm}
    \subfigure{\includegraphics[width=0.28\linewidth, height=0.18\hsize]{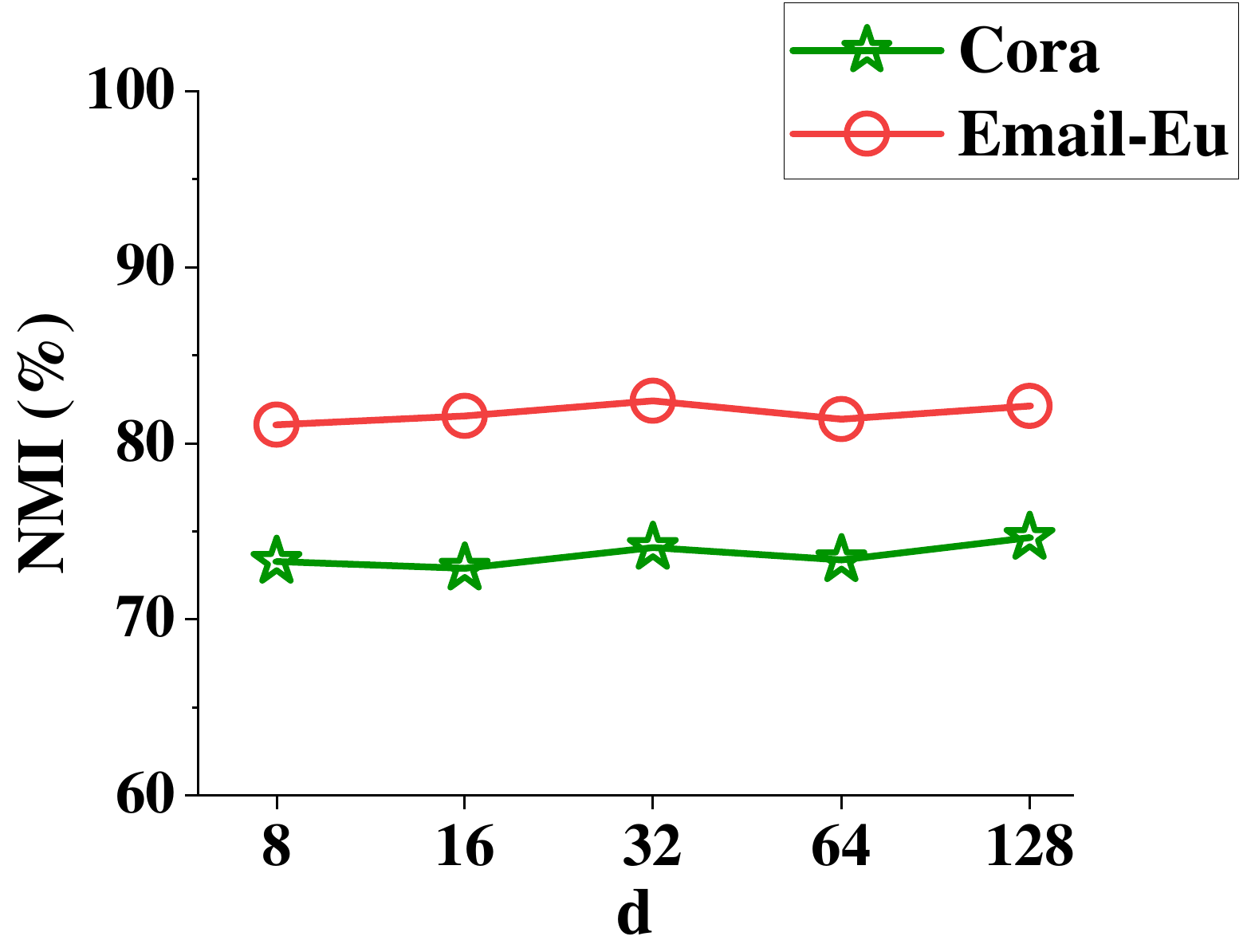}}\hspace{0.5cm}
    \subfigure{\includegraphics[width=0.28\linewidth, height=0.18\hsize]{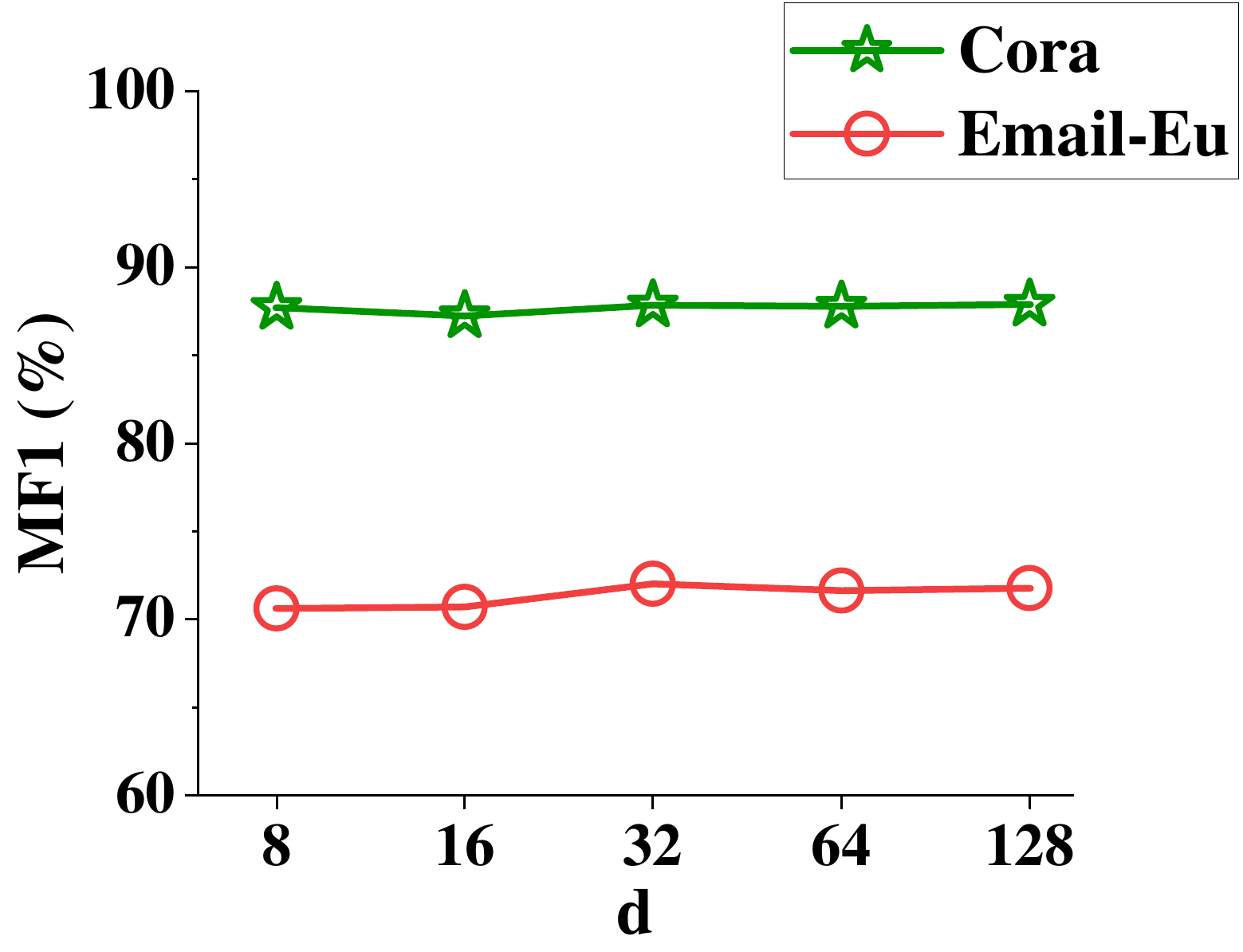}}\par
    \vspace{-2ex}
    \subfigure{\includegraphics[width=0.28\linewidth, height=0.18\hsize]{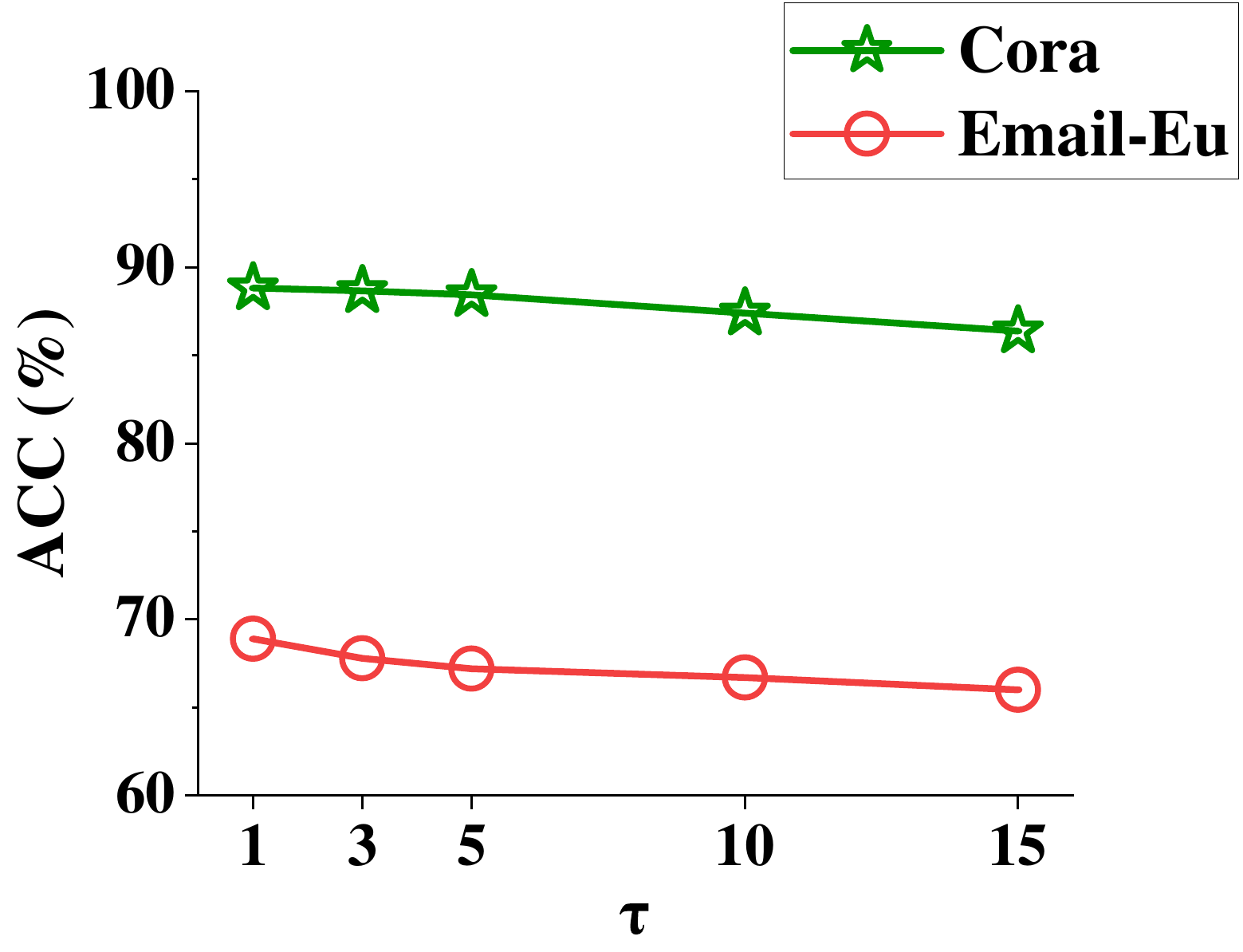}}\hspace{0.5cm}
    \subfigure{\includegraphics[width=0.28\linewidth, height=0.18\hsize]{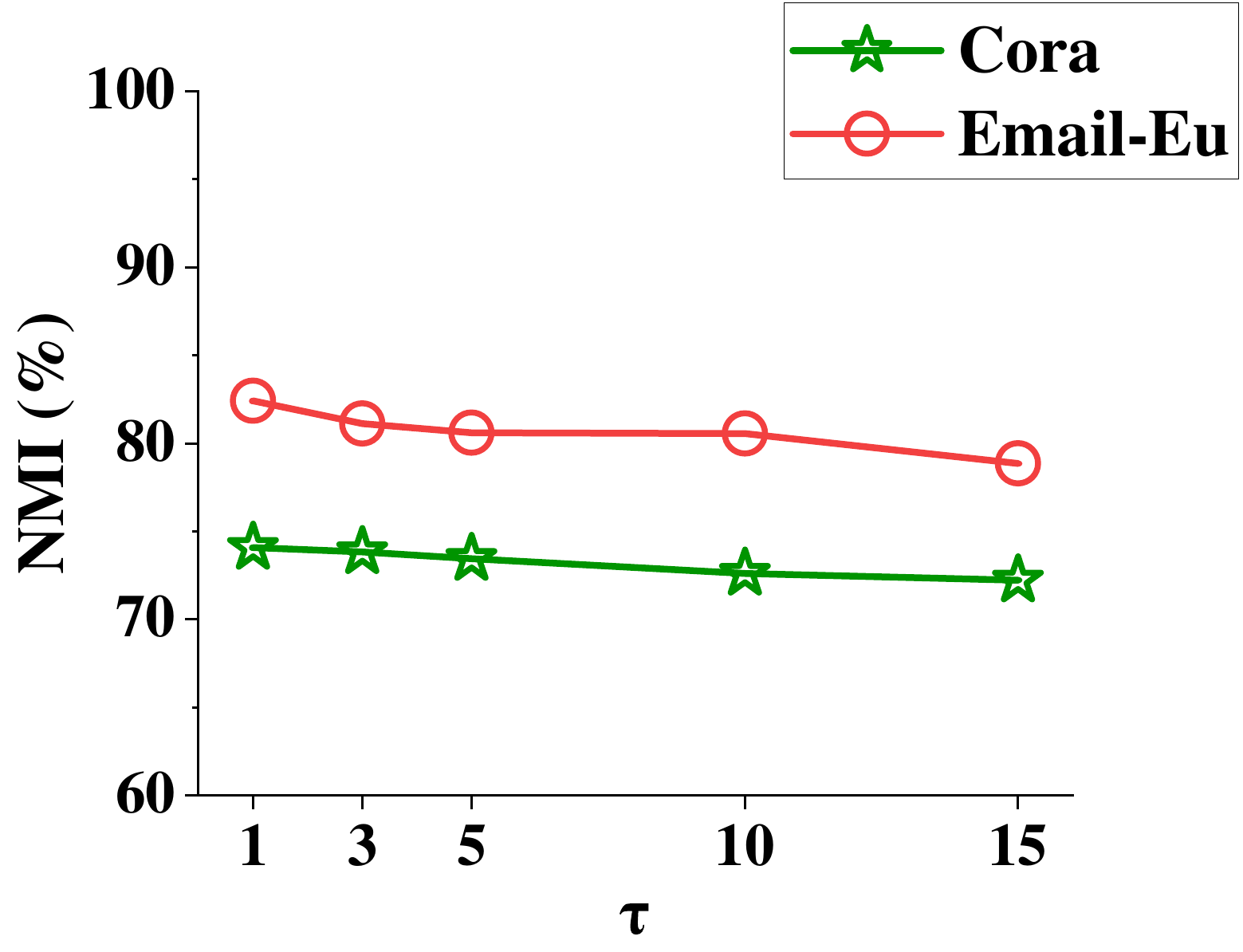}}\hspace{0.5cm}
    \subfigure{\includegraphics[width=0.28\linewidth, height=0.18\hsize]{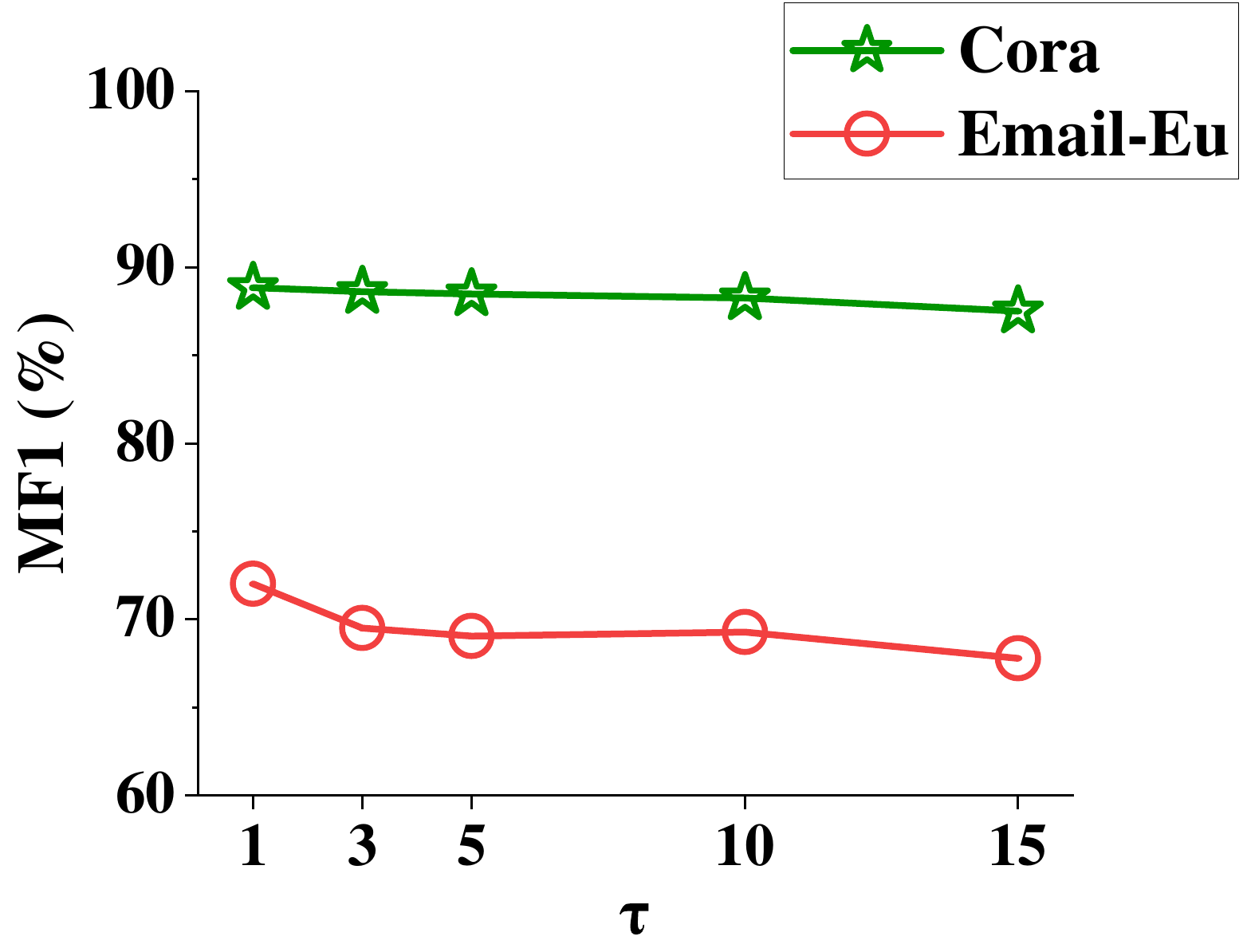}}\par
    \caption{Sensitivity analysis of hyperparameters ($d, \tau$).}
    \label{fig:Hyperparameter}
\end{figure*}

\subsection{The HGP Algorithm Evaluation -- RQ3}
We evaluate the performance of the GCLS$^2$ combined with the HGP algorithm on large-scale datasets and compare the HGP algorithm with state-of-the-art graph partitioning algorithms, including METIS, LDG, and Hash.
As shown in Table~\ref{tab:large-p}, the GCLS$^2$ with HGP achieves the best performance on the TWeibo and AmazonProducts datasets, outperforming the second-best by 4\% in MF1. Note that ``OOM'' denotes out-of-memory errors.
Figure~\ref{fig:hgp-l} shows that the maximum load difference of LDG on TWeibo and AmazonProducts is close to $10^6$, illustrating the inability of the LDG algorithm to maintain good load balancing.
This makes it challenging for small-memory machines to support online training.

Figure~\ref{fig:hdg} presents the performance evaluation of the HGP algorithm.
It can be seen that the HGP algorithm can partition hundreds of millions of edge-level graphs in approximately $10^3$ seconds while ensuring load balancing across 256 partitions.
In contrast, the Hash algorithm completes partitioning in less than 1 second by performing only a random hash computation for each node.
Moreover, as seen in Figure~\ref{fig:hgp-l}, the HGP algorithm achieves good load balancing, reducing the maximum load difference by one order of magnitude compared to METIS and three orders of magnitude compared to LDG.
From Figure~\ref{fig:hgp-p}, when the partition size $p=32$, the HGP algorithm completes the partitioning of graphs with tens of millions of edges in just 155.91 seconds.
Combining these findings with Table~\ref{tab:large-d}, it is clear that the GCLS$^2$+HGP method performs well in terms of efficiency and accuracy of community detection in large-scale datasets.

\begin{table}[!t]
    \centering
    \caption{Performance of GCLS$^2$ with different graph partitioning algorithms on large-scale graph datasets.}
    \label{tab:large-p}
    \resizebox{\linewidth}{!}{
    \begin{tabular}{cccccccc}
        \toprule
        \multirow{2}{*}[-0.5ex]{\bf Methods} & \multicolumn{3}{c}{\bf TWeibo} & \phantom{a}
        & \multicolumn{3}{c}{\bf AmazonProducts}\\
        \cmidrule{2-4} \cmidrule{6-8}
        & ACC &  NMI  & MF1 & & ACC & NMI & MF1\\
        \midrule
        {\bf \shortstack{GCLS$^2$+METIS}} & \underline{41.83} & \underline{42.90} & \underline{55.31} & & \underline{58.73} & \underline{51.82} & \underline{59.40}\\
        {\bf \shortstack{GCLS$^2$+LDG}} & OOM & OOM & OOM & & OOM & OOM & OOM\\
        {\bf \shortstack{GCLS$^2$+Hash}} & 38.62 & 35.33 & 40.15 & & 55.52 & 49.29 & 57.35\\
        \midrule
        {\bf \shortstack{GCLS$^2$+HGP}} & \textbf{42.94} & \textbf{44.72} & \textbf{59.92} & & \textbf{61.79} & \textbf{53.01} & \textbf{63.54}\\
        \bottomrule
    \end{tabular}
    }
\end{table}

\begin{figure*}[!ht]
    \centering
    \subfigure[$\varepsilon$=10]{\includegraphics[width=0.27\linewidth, height=0.19\hsize]{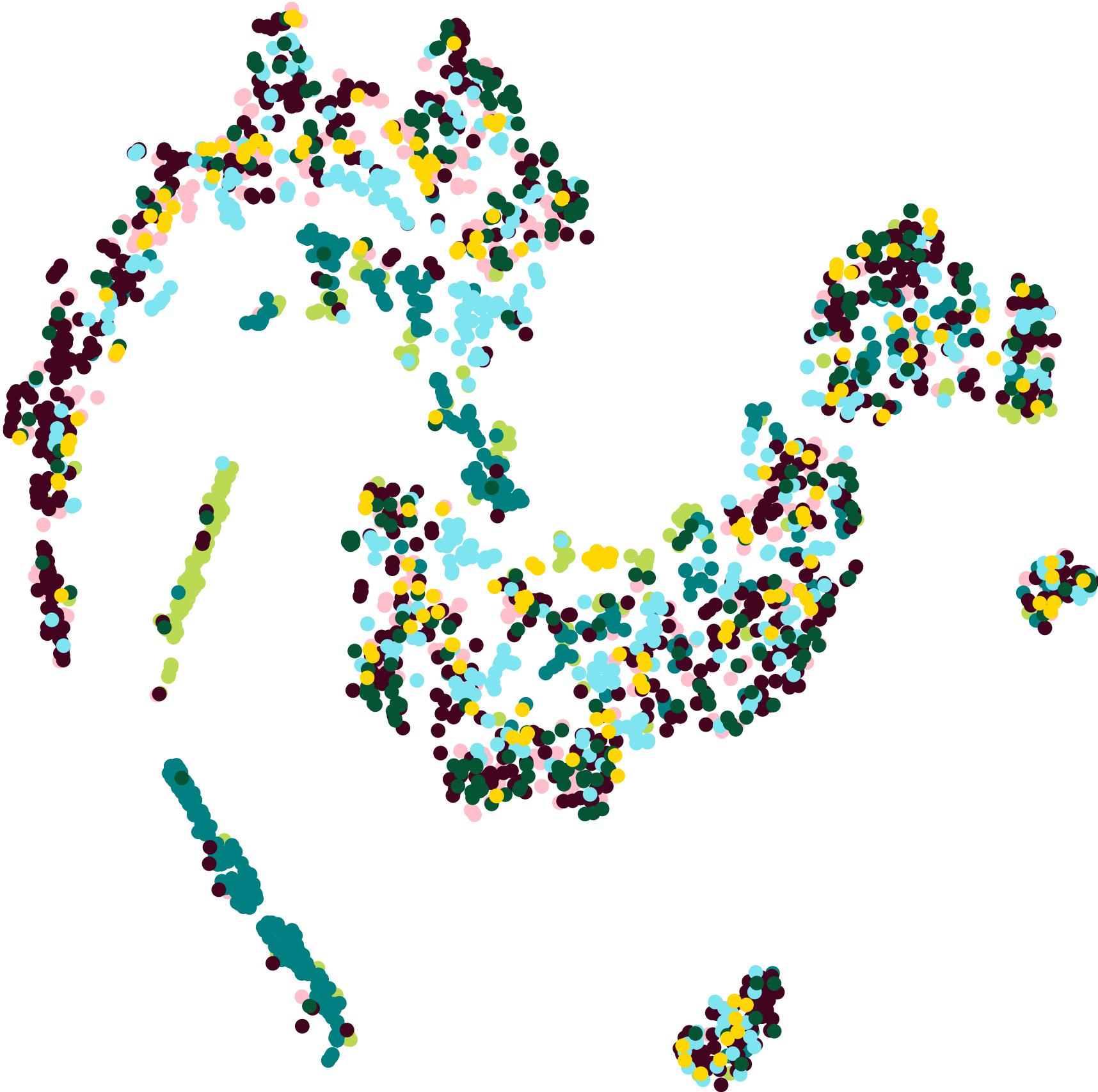}}\hspace{1cm}
    \subfigure[$\varepsilon$=30]{\includegraphics[width=0.27\linewidth, height=0.19\hsize]{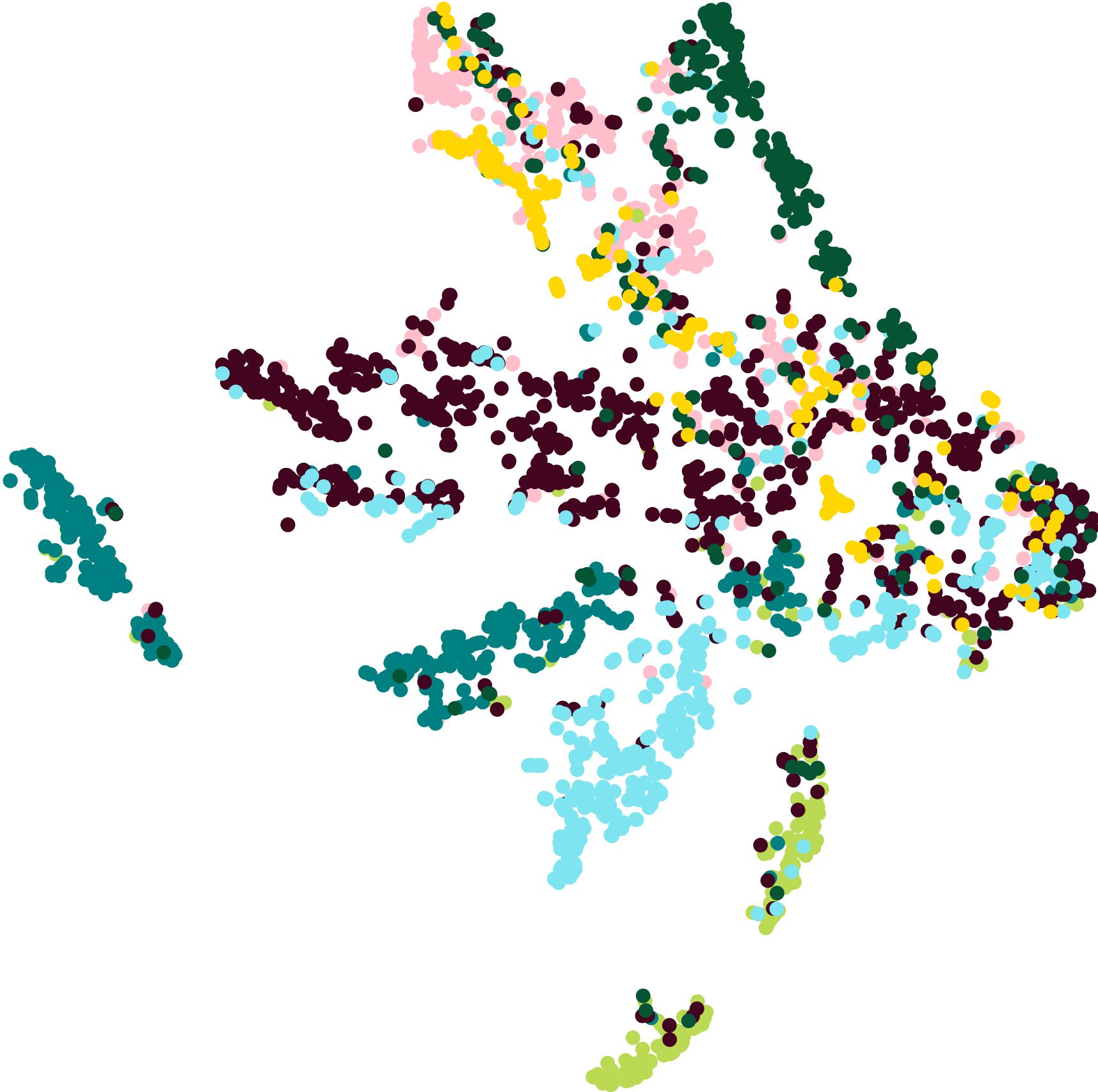}}\hspace{1cm}
    \subfigure[$\varepsilon$=500]{\includegraphics[width=0.27\linewidth, height=0.19\hsize]{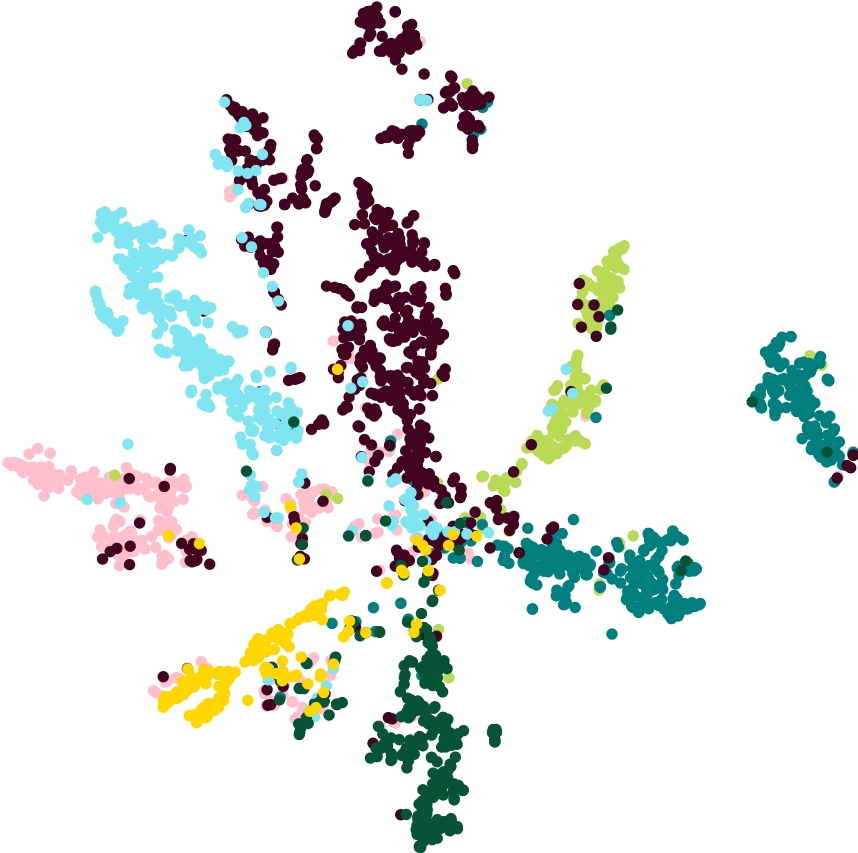}}\par
    \caption{Community detection performance on Cora.}
    \label{fig:subfigures}
\end{figure*}

\subsection{The Hyperparameter Analysis -- RQ4}
Figure~\ref{fig:Hyperparameter} shows the sensitivity analysis of the hyperparameters (structure semantic dimension $d$ and temperature parameter $\tau$) of GCLS$^2$ on Cora and Email-Eu datasets. Then we can answer RQ4.
From the results, the ACC, NMI, and MF1 metrics are stable on Cora and Email-Eu datasets for $d$=\{8, 16, 32, 64, 128\}. 
However, the temperature parameter $\tau$ has the best performance when the $\tau$=1.
Since the $\tau$ affects the vector distance of the sample pairs for structure contrastive learning, a larger $\tau$ reduces the effect of structure contrastive learning, resulting in the reduction of ACC, NMI, and MF1 metrics for both datasets.



\subsection{The Case Study}
We evaluate the community detection performance of the GCLS$^2$ model on Cora by the T-SNE~\cite{van2008visualizing} downscaling. 
Specifically, after the training of Algorithm~\ref{alg:train} for $\varepsilon$ = \{10, 30, 500\}, respectively, we visualize the distribution of node features $Z$ in two dimensions, using nodes in 7 different colors to represent the 7 communities.
The results are shown in Figure~\ref{fig:subfigures}. 
We can clearly see that from $\varepsilon$ = 10 to 30 iterations, the labeled nodes are closing in on the communities corresponding to the labels. By 500 iterations, the detection accuracy is significantly improved, validating the effectiveness of our GCLS$^2$ model.

\section{Related Work}
\label{sec:RW}
\noindent \textbf{Community Detection:} The community detection can be classified into traditional and deep learning methods. The former one mainly explores communities from network structure with clustering and optimization algorithms~\cite{lancichinetti2009community, jian2018efficiently}, whereas the latter one utilizes deep learning to uncover deep network information and model complex relationships from high-dimensional attribute data to lower-dimensional vectors. 
Due to complex topology structures and attribute features for real-world networks, high computational costs make traditional methods less applicable to practical applications. 
However, existing deep learning methods (e.g., GCN~\cite{kipf2016semi} and GAT~\cite{velivckovic2018graph}) rely on real labels and attributes of the data graph and pay little attention to the important structure of communities.
In contrast, our GCLS$^2$ method starts from unsupervised learning that does not rely on real labels and attributes. Moreover, GCLS$^2$ performs effective community detection by mining the structure information of the community.

\noindent \textbf{Graph Contrastive Learning:} The graph contrastive learning optimizes the representation of nodes, by maximizing the agreement between pairs of positive samples and minimizing the agreement between pairs of negative samples to be adapted to various downstream tasks.
For the node level, there are two general contrastive ways: 
\romannumeral1) node-graph, for example, DGI~\cite{velivckovicdeep} contrasts node feature embeddings of augmented view and original data graph with graph feature embeddings of the original data graph, and MVGRL~\cite{hassani2020contrastive} contrasts node feature embeddings of one view with the graph feature embeddings of another view, \romannumeral2) node-node, for example, GRACE~\cite{zhu2020deep} contrasts node feature embeddings of two augmented views with each other, and NCLA~\cite{shen2023neighbor} contrasts neighbors' feature embeddings of multiple augmented views.
Although these contrastive learning methods have made greater progress in the feature representation, most contrasted samples in the community detection task are against the community's inherent information representation, where feature embedding representations within communities should be similar and feature embedding representations between communities should be dissimilar.
Our proposed GCLS$^2$ approach uses a high-level structure adjacency matrix as a signal to guide the anchor closer to dense intra-community and away from the inter-community and achieves good detection results even when using community structure information only.
 
To the best of our knowledge, no prior works proposed to use the structure contrastive loss for the community detection task. 
The ability to operate without relying on real labels and attributes makes our GCLS$^2$ model highly versatile for real-world applications.
Moreover, we present our theoretical analysis for structural contrast loss training.
We also design a graph partitioning algorithm to scale our proposed approach to handle large-scale graphs 
and training can be accelerated by parallel processing.



\section{Conclusion}
\label{sec:CC}
In this paper, we analyze the limitations of traditional GCL methods for community detection, and propose a novel Graph Contrastive Learning with Structure Semantics (CGLS$^2$) framework for the community detection.
We construct a high-level structure view based on classical community structure, and then extract structural and attribute semantics through the structure similarity semantics encoder to obtain a comprehensive node feature representation.
We also design the structure contrastive learning to enhance the structural feature representation of nodes for more accurate community detection.
Moreover, we propose the HGP algorithm, which makes it possible to perform online training and community detection over large-scale graph datasets on low-memory machines.
Extensive experiments confirm the effectiveness of our GCLS$^2$ approach and HGP algorithm for the community detection and large-scale graph partitioning.


\clearpage
\bibliography{db,IEEEabrv}

\begin{thebibliography}{10}
\providecommand{\url}[1]{#1}
\csname url@samestyle\endcsname
\providecommand{\newblock}{\relax}
\providecommand{\bibinfo}[2]{#2}
\providecommand{\BIBentrySTDinterwordspacing}{\spaceskip=0pt\relax}
\providecommand{\BIBentryALTinterwordstretchfactor}{4}
\providecommand{\BIBentryALTinterwordspacing}{\spaceskip=\fontdimen2\font plus
\BIBentryALTinterwordstretchfactor\fontdimen3\font minus \fontdimen4\font\relax}
\providecommand{\BIBforeignlanguage}[2]{{%
\expandafter\ifx\csname l@#1\endcsname\relax
\typeout{** WARNING: IEEEtran.bst: No hyphenation pattern has been}%
\typeout{** loaded for the language `#1'. Using the pattern for}%
\typeout{** the default language instead.}%
\else
\language=\csname l@#1\endcsname
\fi
#2}}
\providecommand{\BIBdecl}{\relax}
\BIBdecl

\bibitem{whisstock2003prediction}
J.~C. Whisstock and A.~M. Lesk, ``Prediction of protein function from protein sequence and structure,'' \emph{Quarterly Reviews of Biophysics}, vol.~36, no.~3, pp. 307--340, 2003.

\bibitem{bedi2016community}
P.~Bedi and C.~Sharma, ``Community detection in social networks,'' \emph{Wiley Interdisciplinary Reviews: Data Mining and Knowledge Discovery}, vol.~6, no.~3, pp. 115--135, 2016.

\bibitem{LiuX0ZHPNYY20}
F.~Liu, S.~Xue, J.~Wu, C.~Zhou, W.~Hu, C.~Paris, S.~Nepal, J.~Yang, and P.~S. Yu, ``Deep learning for community detection: Progress, challenges and opportunities,'' in \emph{Proceedings of the International Joint Conference on Artificial Intelligence, (IJCAI)}, 2020, pp. 4981--4987.

\bibitem{keyvanpour2020ad}
M.~R. Keyvanpour, M.~B. Shirzad, and M.~Ghaderi, ``Ad-c: A new node anomaly detection based on community detection in social networks,'' \emph{International Journal of Electronic Business}, vol.~15, no.~3, pp. 199--222, 2020.

\bibitem{zarandi2018community}
F.~D. Zarandi and M.~K. Rafsanjani, ``Community detection in complex networks using structural similarity,'' \emph{Physica A: Statistical Mechanics and its Applications}, vol. 503, pp. 882--891, 2018.

\bibitem{amini2013pseudo}
A.~A. Amini, A.~Chen, P.~J. Bickel, and E.~Levina, ``Pseudo-likelihood methods for community detection in large sparse networks,'' \emph{The Annals of Statistics}, pp. 2097--2122, 2013.

\bibitem{li2016multi}
Z.~Li and J.~Liu, ``A multi-agent genetic algorithm for community detection in complex networks,'' \emph{Physica A: Statistical Mechanics and its Applications}, vol. 449, pp. 336--347, 2016.

\bibitem{ChenLB19}
Z.~Chen, L.~Li, and J.~Bruna, ``Supervised community detection with line graph neural networks,'' in \emph{International Conference on Learning Representations (ICLR)}, 2019.

\bibitem{xin2017deep}
X.~Xin, C.~Wang, X.~Ying, and B.~Wang, ``Deep community detection in topologically incomplete networks,'' \emph{Physica A: Statistical Mechanics and its Applications}, vol. 469, pp. 342--352, 2017.

\bibitem{he2020momentum}
K.~He, H.~Fan, Y.~Wu, S.~Xie, and R.~Girshick, ``Momentum contrast for unsupervised visual representation learning,'' in \emph{Proceedings of the IEEE/CVF Conference on Computer Vision and Pattern Recognition}, 2020, pp. 9729--9738.

\bibitem{zhu2020deep}
Y.~Zhu, Y.~Xu, F.~Yu, Q.~Liu, S.~Wu, and L.~Wang, ``{Deep Graph Contrastive Representation Learning},'' in \emph{ICML Workshop on Graph Representation Learning and Beyond}, 2020.

\bibitem{zhu2021graph}
{Zhu, Yanqiao and Xu, Yichen and Yu, Feng and Liu, Qiang and Wu, Shu and Wang, Liang}, ``Graph contrastive learning with adaptive augmentation,'' in \emph{Proceedings of the Web Conference (WWW)}, 2021, pp. 2069--2080.

\bibitem{kong2019k}
Y.-X. Kong, G.-Y. Shi, R.-J. Wu, and Y.-C. Zhang, ``k-core: Theories and applications,'' \emph{Physics Reports}, vol. 832, pp. 1--32, 2019.

\bibitem{huang2014querying}
X.~Huang, H.~Cheng, L.~Qin, W.~Tian, and J.~X. Yu, ``Querying k-truss community in large and dynamic graphs,'' in \emph{Proceedings of the ACM SIGMOD International Conference on Management of Data (SIGMOD)}, 2014, pp. 1311--1322.

\bibitem{gregori2012parallel}
E.~Gregori, L.~Lenzini, and S.~Mainardi, ``Parallel k-clique community detection on large-scale networks,'' \emph{IEEE Transactions on Parallel and Distributed Systems}, vol.~24, no.~8, pp. 1651--1660, 2012.

\bibitem{zhang2023top}
Y.~Ye, X.~Lian, and M.~Chen, ``Efficient exact subgraph matching via gnn-based path dominance embedding,'' \emph{Proceedings of the VLDB Endowment}, vol.~17, no.~7, 2024.

\bibitem{rai2023top}
N.~Rai and X.~Lian, ``Top-$ k $ community similarity search over large-scale road networks,'' \emph{IEEE Transactions on Knowledge and Data Engineering}, vol.~35, no.~10, pp. 10\,710--10\,721, 2023.

\bibitem{frith2020minimally}
M.~C. Frith, L.~No{\'e}, and G.~Kucherov, ``Minimally overlapping words for sequence similarity search,'' \emph{Bioinformatics}, vol.~36, no. 22-23, pp. 5344--5350, 2020.

\bibitem{zhao2012rapsearch2}
Y.~Zhao, H.~Tang, and Y.~Ye, ``Rapsearch2: a fast and memory-efficient protein similarity search tool for next-generation sequencing data,'' \emph{Bioinformatics}, vol.~28, no.~1, pp. 125--126, 2012.

\bibitem{yang2012defining}
J.~Yang and J.~Leskovec, ``Defining and evaluating network communities based on ground-truth,'' in \emph{Proceedings of the ACM SIGKDD Workshop on Mining Data Semantics}, 2012, pp. 1--8.

\bibitem{chang2022efficient}
L.~Chang, M.~Xu, and D.~Strash, ``Efficient maximum k-plex computation over large sparse graphs,'' \emph{Proceedings of the VLDB Endowment}, vol.~16, no.~2, pp. 127--139, 2022.

\bibitem{karypis1998fast}
G.~Karypis and V.~Kumar, ``A fast and high quality multilevel scheme for partitioning irregular graphs,'' \emph{SIAM Journal on scientific Computing}, vol.~20, no.~1, pp. 359--392, 1998.

\bibitem{stanton2012streaming}
I.~Stanton and G.~Kliot, ``Streaming graph partitioning for large distributed graphs,'' in \emph{Proceedings of the 18th ACM SIGKDD international conference on Knowledge discovery and data mining}, 2012, pp. 1222--1230.

\bibitem{tsourakakis2014fennel}
C.~Tsourakakis, C.~Gkantsidis, B.~Radunovic, and M.~Vojnovic, ``Fennel: Streaming graph partitioning for massive scale graphs,'' in \emph{Proceedings of the 7th ACM international conference on Web search and data mining}, 2014, pp. 333--342.

\bibitem{xie2014distributed}
C.~Xie, L.~Yan, W.-J. Li, and Z.~Zhang, ``Distributed power-law graph computing: Theoretical and empirical analysis,'' \emph{Advances in neural information processing systems}, vol.~27, 2014.

\bibitem{petroni2015hdrf}
F.~Petroni, L.~Querzoni, K.~Daudjee, S.~Kamali, and G.~Iacoboni, ``Hdrf: Stream-based partitioning for power-law graphs,'' in \emph{Proceedings of the 24th ACM international on conference on information and knowledge management}, 2015, pp. 243--252.

\bibitem{zhang2017graph}
C.~Zhang, F.~Wei, Q.~Liu, Z.~G. Tang, and Z.~Li, ``Graph edge partitioning via neighborhood heuristic,'' in \emph{Proceedings of the 23rd ACM SIGKDD International Conference on Knowledge Discovery and Data Mining}, 2017, pp. 605--614.

\bibitem{bachman2019learning}
P.~Bachman, R.~D. Hjelm, and W.~Buchwalter, ``Learning representations by maximizing mutual information across views,'' \emph{Advances in neural information processing systems}, vol.~32, 2019.

\bibitem{tschannenmutual}
M.~Tschannen, J.~Djolonga, P.~K. Rubenstein, S.~Gelly, and M.~Lucic, ``On mutual information maximization for representation learning,'' in \emph{International Conference on Learning Representations (ICLR)}.

\bibitem{poole2019variational}
B.~Poole, S.~Ozair, A.~Van Den~Oord, A.~Alemi, and G.~Tucker, ``On variational bounds of mutual information,'' in \emph{International Conference on Machine Learning}.\hskip 1em plus 0.5em minus 0.4em\relax PMLR, 2019, pp. 5171--5180.

\bibitem{oord2018representation}
A.~v.~d. Oord, Y.~Li, and O.~Vinyals, ``Representation learning with contrastive predictive coding,'' \emph{arXiv preprint arXiv:1807.03748}, 2018.

\bibitem{gong2023ma}
X.~Gong, C.~Yang, and C.~Shi, ``Ma-gcl: Model augmentation tricks for graph contrastive learning,'' in \emph{Proceedings of the AAAI Conference on Artificial Intelligence (AAAI)}, 2023, pp. 4284--4292.

\bibitem{hu2020gpt}
Z.~Hu, Y.~Dong, K.~Wang, K.-W. Chang, and Y.~Sun, ``Gpt-gnn: Generative pre-training of graph neural networks,'' in \emph{Proceedings of the 26th ACM SIGKDD international conference on knowledge discovery \& data mining}, 2020, pp. 1857--1867.

\bibitem{shen2023neighbor}
X.~Shen, D.~Sun, S.~Pan, X.~Zhou, and L.~T. Yang, ``Neighbor contrastive learning on learnable graph augmentation,'' in \emph{Proceedings of the AAAI Conference on Artificial Intelligence (AAAI)}, 2023, pp. 9782--9791.

\bibitem{chen2023attribute}
J.~Chen and G.~Kou, ``Attribute and structure preserving graph contrastive learning,'' in \emph{Proceedings of the AAAI Conference on Artificial Intelligence (AAAI)}, 2023, pp. 7024--7032.

\bibitem{sen2008collective}
P.~Sen, G.~Namata, M.~Bilgic, L.~Getoor, B.~Galligher, and T.~Eliassi-Rad, ``Collective classification in network data,'' \emph{AI Magazine}, vol.~29, no.~3, pp. 93--93, 2008.

\bibitem{shchur2018pitfalls}
O.~Shchur, M.~Mumme, A.~Bojchevski, and S.~G{\"u}nnemann, ``Pitfalls of graph neural network evaluation,'' \emph{arXiv preprint arXiv:1811.05868}, 2018.

\bibitem{YinBLG17}
H.~Yin, A.~R. Benson, J.~Leskovec, and D.~F. Gleich, ``Local higher-order graph clustering,'' in \emph{Proceedings of the International Conference on Knowledge Discovery and Data Mining (SIGKDD)}.\hskip 1em plus 0.5em minus 0.4em\relax ACM, 2017, pp. 555--564.

\bibitem{yang2023pane}
R.~Yang, J.~Shi, X.~Xiao, Y.~Yang, S.~S. Bhowmick, and J.~Liu, ``Pane: scalable and effective attributed network embedding,'' \emph{The VLDB Journal}, vol.~32, no.~6, pp. 1237--1262, 2023.

\bibitem{GraphSAINT}
H.~Zeng, H.~Zhou, A.~Srivastava, R.~Kannan, and V.~K. Prasanna, ``Graphsaint: Graph sampling based inductive learning method,'' in \emph{8th International Conference on Learning Representations, {ICLR} 2020, Addis Ababa, Ethiopia, April 26-30, 2020}, 2020.

\bibitem{kipf2016semi}
T.~N. Kipf and M.~Welling, ``Semi-supervised classification with graph convolutional networks,'' in \emph{International Conference on Learning Representations (ICLR)}, 2017.

\bibitem{gat}
P.~Velickovic, G.~Cucurull, A.~Casanova, A.~Romero, P.~Li{\`{o}}, and Y.~Bengio, ``Graph attention networks,'' in \emph{6th International Conference on Learning Representations, {ICLR} 2018, Vancouver, BC, Canada, April 30 - May 3, 2018, Conference Track Proceedings}, 2018.

\bibitem{velivckovicdeep}
P.~Veli{\v{c}}kovi{\'c}, W.~Fedus, W.~L. Hamilton, P.~Li{\`o}, Y.~Bengio, and R.~D. Hjelm, ``Deep graph infomax,'' in \emph{International Conference on Learning Representations (ICLR)}, 2018.

\bibitem{hassani2020contrastive}
K.~Hassani and A.~H. Khasahmadi, ``Contrastive multi-view representation learning on graphs,'' in \emph{International Conference on Machine Learning (ICML)}.\hskip 1em plus 0.5em minus 0.4em\relax PMLR, 2020, pp. 4116--4126.

\bibitem{mo2022simple}
Y.~Mo, L.~Peng, J.~Xu, X.~Shi, and X.~Zhu, ``Simple unsupervised graph representation learning,'' in \emph{Proceedings of the AAAI Conference on Artificial Intelligence (AAAI)}, 2022, pp. 7797--7805.

\bibitem{abbas2018streaming}
Z.~Abbas, V.~Kalavri, P.~Carbone, and V.~Vlassov, ``Streaming graph partitioning: an experimental study,'' \emph{Proceedings of the VLDB Endowment}, vol.~11, no.~11, pp. 1590--1603, 2018.

\bibitem{kong2011robust}
D.~Kong, C.~Ding, and H.~Huang, ``Robust nonnegative matrix factorization using l21-norm,'' in \emph{Proceedings of the ACM International Conference on Information and Knowledge Management}, 2011, pp. 673--682.

\bibitem{van2008visualizing}
L.~Van~der Maaten and G.~Hinton, ``Visualizing data using t-sne.'' \emph{Journal of Machine Learning Research}, vol.~9, no.~11, 2008.

\bibitem{lancichinetti2009community}
A.~Lancichinetti and S.~Fortunato, ``Community detection algorithms: a comparative analysis,'' \emph{Physical Review E—Statistical, Nonlinear, and Soft Matter Physics}, vol.~80, no.~5, p. 056117, 2009.

\bibitem{jian2018efficiently}
X.~Jian, X.~Lian, and L.~Chen, ``On efficiently detecting overlapping communities over distributed dynamic graphs,'' in \emph{International Conference on Data Engineering (ICDE)}.\hskip 1em plus 0.5em minus 0.4em\relax IEEE, 2018, pp. 1328--1331.

\bibitem{velivckovic2018graph}
P.~Veli{\v{c}}kovi{\'c}, G.~Cucurull, A.~Casanova, A.~Romero, P.~Li{\`o}, and Y.~Bengio, ``Graph attention networks,'' in \emph{International Conference on Learning Representations (ICLR)}, 2018.

\end{thebibliography}
\bibliographystyle{IEEEtran}

\end{document}